\documentclass[reprint,aps,prl,amsmath,amssymb,superscriptaddress]{revtex4-2}
\usepackage[T1]{fontenc}
\usepackage[utf8]{inputenc}

\usepackage{newtxtext}
\usepackage{newtxmath}

\usepackage{soul}
\usepackage{graphicx}
\usepackage{dcolumn}
\usepackage{bm}
\usepackage{tikz}
\usepackage{newtxtext}
\usepackage{mathtools}
\DeclareMathOperator*{\argmin}{arg\,min}
\DeclareMathOperator*{\argmax}{arg\,max}
\usepackage{hyperref}
\usepackage{xcolor}
\definecolor{mydarkred}{RGB}{139,0,0}
\definecolor{mydarkred2}{RGB}{199,0,0}
\hypersetup{colorlinks=true, linkcolor=mydarkred, citecolor=mydarkred, urlcolor=mydarkred}

\newcommand{\PRLsection}[1]{\emph{#1}.---}

\begin{document}

\newcommand{\corrNote}{Corresponding authors: \href{mailto:artemyk@gmail.com}{artemyk@gmail.com}, \href{mailto:maguilera@bcamath.org}{maguilera@bcamath.org}. These authors contributed equally to this work.}

\title{Inferring entropy production in many-body systems using nonequilibrium maximum entropy}
\author{Miguel Aguilera}
\altaffiliation{\corrNote}
\affiliation{BCAM -- Basque Center for Applied Mathematics, 48009 Bilbao, Spain}
\affiliation{IKERBASQUE, Basque Foundation for Science, 48009 Bilbao, Spain}
\author{Sosuke Ito}
\affiliation{Universal Biology Institute, The University of Tokyo, 7-3-1 Hongo,
Bunkyo-ku, Tokyo 113-0033, Japan}
\affiliation{Department of Physics, The University of Tokyo, 7-3-1 Hongo, Bunkyo-ku,
Tokyo 113-0033, Japan}
\author{Artemy Kolchinsky}
\altaffiliation{\corrNote}
\affiliation{ICREA-Complex Systems Lab, Universitat Pompeu Fabra, 08003 Barcelona,
Spain}
\affiliation{Universal Biology Institute, The University of Tokyo, 7-3-1 Hongo,
Bunkyo-ku, Tokyo 113-0033, Japan}
\begin{abstract}
We propose a method for inferring entropy production (EP) in high-dimensional stochastic systems, including many-body systems and non-Markovian systems with long memory. Standard techniques for estimating EP become intractable in such systems due to computational and statistical limitations. We infer trajectory-level EP and lower bounds on average EP by exploiting a nonequilibrium analogue of the Maximum Entropy principle, along with convex duality. Our approach uses only samples of trajectory observables, such as spatiotemporal correlations. It does not require reconstruction of high-dimensional probability distributions or rate matrices, nor impose any special assumptions such as discrete states or multipartite dynamics. In addition, it may be used to compute a hierarchical decomposition of EP, reflecting contributions from different interaction orders, and it has an intuitive physical interpretation as a ``thermodynamic uncertainty relation.'' We demonstrate its numerical performance on a disordered nonequilibrium spin model with 1000 spins and a large neural spike-train dataset.
\end{abstract}

\maketitle

\newcommand\gobs{\bm{g}}
\newcommand\condpi{p_i}
\newcommand\condpirev{\tilde{p}_i}
\newcommand\condqi{q_i}
\newcommand\condpiZ{p_0}
\newcommand\condpirevZ{\tilde{p}_0}
\newcommand\condqiZ{q_0}
\global\long\def\dims{N}
\global\long\def\ntimes{T}
\newcommand{\cev}[1]{\reflectbox{\ensuremath{\vec{\reflectbox{\ensuremath{#1}}}}}}
\def\params{{\bm{\theta}}}
\def\xx{{\bm{x}}}
\def\traj{{\xx}}
\def\revtraj{\tilde\xx}
\def\ep{{\Sigma}}
\def\epG{\ep_{\gobs}}
\def\epGi{\epG^{(i)}}
\def\GepG{\widehat{\ep}_{\gobs}}
\def\epGL{\ep_{\gobs}^\textrm{L}}
\def\epGtur{\ep_{\gobs}^\textrm{TUR}}
\def\epNotG{\ep^{\perp}_{\gobs}}
\def\epR{\ep_2}
\def\pst{p^{\rm st}}
\newcommand\K{\mathsf{K}_{\tilde p}}
\newcommand\Kinv{\K^{-1}}
\def\pr#1{\left( #1\right)}
\def\ang#1{\left\langle #1\right\rangle}
\def\rang#1{\langle #1\rangle}
\def\Gvar{G}
\def\Trajvar{X}
\def\epGvar{\epG^\mathrm{DPI}}

\newcommand\nnfunc{v}
\newcommand\epMine{\Sigma_{\gobs}^{\text{B}}}
\newcommand\epKim{\Sigma_{\gobs}^{\text{KO}}}

\newcommand\eptraj{\sigma_{\params^*}}

The central quantity of interest in nonequilibrium thermodynamics is \emph{entropy production} (EP)~\cite{seifert2012stochastic}. In microscopic physical systems, EP characterizes departure from thermodynamic equilibrium and quantifies the dissipation of thermodynamic free energy. More generally, EP provides an information-theoretic measure of temporal irreversibility, including in meso- and macroscopic systems~\cite{gallavotti1995dynamical,kurchan1998fluctuation,maes1999fluctuation,jarzynski2000hamiltonian}.

Recently, there has been growing interest in \emph{thermodynamic inference}~\cite{seifert2019stochastic}, that is, the problem of estimating EP from empirical measurements of a stochastic system. Various methods have been developed for inferring EP from partial~\cite{rahav2007fluctuation,esposito2012stochastic,bo2014entropy,wang2016entropy,bisker2017hierarchical,polettini2017effective,ehrich2021tightest,nitzan2023universal} information. For instance, the celebrated \emph{thermodynamic uncertainty relation} (TUR)~\cite{barato2015thermodynamic,gingrich2016dissipation,pietzonka2017finite,manikandan2020inferring} bounds EP in terms of the mean and variance of a single current. Other techniques relate EP to the statistics of waiting times~\cite{skinner2021estimating,van2022thermodynamic,harunari2022learn,martinez2019inferring}, observed transitions~\cite{skinner2021improved,van2023time,bisker2017hierarchical,polettini2017effective}, and counting observables~\cite{pietzonka2024thermodynamic}. Many of these methods are designed to infer underlying dissipation from a small number of coarse-grained observables.

In this Letter, we consider thermodynamic inference in the high-dimensional setting, where a large number of observables are available. This is particularly relevant for multivariate measurements of complex systems --- such as nonequilibrium disordered networks~\cite{aguilera2023nonequilibrium}, biological active matter~\cite{battle2016broken,tan2022odd}, and neural systems~\cite{lynn2021broken,de2023temporal,sekizawa2024decomposing,geli2025non,nartallo2025nonequilibrium} --- that have many degrees of freedom and/or long non-Markovian memory.  
A naïve approach requires the estimation of trajectory probability distributions, but in high dimensions, this is usually statistically and numerically infeasible.

To address this challenge, we propose to infer EP using an information-theoretic variational principle, which can be understood as the nonequilibrium analogue of
the Maximum Entropy Principle (MaxEnt) in statistical physics. Our variational principle has a simple physical interpretation: it quantifies the irreversibility captured by the expectations of some (possibly large) number of trajectory observables. 
Importantly, this variational principle has a dual form that leads to %
a tractable %
convex optimization problem. This optimization problem gives a lower bound on EP and an estimate of the fluctuating trajectory-level EP --- directly from trajectory samples and without explicit use of trajectory probabilities. 

As we show, our bound can also be understood as a higher-order thermodynamic uncertainty relation (TUR)~\cite{dechant2020fluctuation,kolchinsky2024generalized}, which we use to derive further simple bounds on EP. Finally, we show that in multipartite systems, our variational principle can be split into smaller subproblems, greatly improving the performance scaling of our method with system size.

\PRLsection{Entropy production} 
We consider a nonequilibrium stochastic system either in continuous or discrete time. 
For time $t\in [0, T]$, the system follows trajectories $\traj$ according to the ``forward'' probability distribution $p(\traj)$. We write $\bm x_t$ for the system's state at time $t$, and $x_{i,t}$ for the state of degree of freedom $i$ at $t$. 
As standard in stochastic thermodynamics, the system is also associated with a ``reverse'' trajectory distribution $\tilde{p}(\traj)$ produced by sampling from the final distribution of the forward process $p(\bm{x}_T)$, applying time-dependent driving in reverse, and finally time-reversing trajectories~\cite{seifert2012stochastic}. 
For simplicity, here we mostly focus on the common case of stationary systems without odd variables (such as velocity). In such system,  
the reverse distribution is given simply by time-reversing forward trajectories, $\tilde p (\traj)=p(\revtraj)$ where $\revtraj_t \vcentcolon= \traj_{T-t}$.

The \emph{trajectory EP} of trajectory $\traj$ and the \emph{average EP} across all trajectories are defined in terms of symmetry breaking between forward and reverse distributions~\cite{seifert2012stochastic},
\begin{align}
    \sigma(\traj) \vcentcolon= \ln \frac{p(\traj)}{\tilde{p}(\traj )}\,,\qquad
    \Sigma \vcentcolon= D(p\Vert\tilde{p})=\ang{ \ln\frac{p(\traj)}{\tilde{p}(\traj )} }_{p} \,,
\label{eq:relent}
\end{align}
where $\langle \ldots\rangle_p= \sum_{\traj}  \ldots p(\traj)$ is the expectation over $p$ and $D(\cdot \Vert \cdot)$ is the Kullback-Leibler (KL) divergence. 
EP %
vanishes in equilibrium, when trajectory statistics are indistinguishable under the forward and reverse distributions.

\PRLsection{EP bound} 
We suppose that one measures trajectory observables, represented by a vector-valued function $\gobs(\traj) \in \mathbb{R}^d$, from the forward and reverse process. Natural choices of observables include correlation functions, e.g., %
two-point ($x_{i,t}x_{j,t'}$) spatiotemporal correlations. We do {not} assume antisymmetric observables such as $\gobs(\traj)=-\gobs(\revtraj)$. 

Our goal is to estimate $\sigma(\traj)$ and $\Sigma$ from forward and reverse samples of observables $\gobs$. The overall irreversibility encoded in these observables is quantified by the KL divergence
between the distributions of $\gobs$ under the forward and reverse processes, $\epGvar \vcentcolon= D(p_{\Gvar}\Vert \tilde p_{\Gvar})$, where $p_{\Gvar} (\gobs')=\ang{\delta_{\gobs',\gobs(\traj)}}_p$ and $\tilde p_{\Gvar}(\gobs')=\ang{\delta_{\gobs',\gobs(\traj)}}_{\tilde p}$. The data processing inequality (DPI) implies that $\epGvar\le \ep$, with equality when $\gobs(\traj)$ is an invertible function~\cite{cover1999elements}.

Estimating $\epGvar$ is very difficult for high-dimensional observables $\gobs$. Instead, we bound EP using a nonequilibrium generalization of the MaxEnt variational principle. 
Specifically, we choose the distribution that minimizes KL divergence to the reverse process while matching the forward-process expectations of the observables: 
\begin{align}
\epG & \vcentcolon= \min_{q}D(q\Vert\tilde{p})\quad\text{subject to} \quad\ang{\gobs}_{q}=\ang{\gobs}_{p}.
\label{eq:me0}
\end{align}
This quantity obeys the bounds $0\le \epG\le \epGvar$, as shown in the \emph{Supplemental Material} (SM) \cite{SupplementalMaterial} \nocite{boyd2004convex,barzilai1988two,lin1999newton}  $\epG$ quantifies the irreversibility captured by observable expectations, and it vanishes if and only if the expectation of $\gobs$ is the same under the forward and reverse processes.
Large-deviations theory gives $\epG$ a physical interpretation in terms of fluctuations: given $n$ sample trajectories from the reverse process, the probability that the empirical average of observables $\gobs$ is equal to $\ang{\gobs}_{p}$ scales as $\asymp e^{-n \epG}$~\cite{touchette2009large}.

Importantly, by exploiting a remarkable information-theoretic duality, we may calculate $\epG$ without optimizing or inferring any probability distributions. As shown in SM~\cite{SupplementalMaterial}, the optimization problem~\eqref{eq:me0} has the dual formulation
\begin{align}
\epG  = \max_{\params \in \mathbb{R}^d} \pr{ \params^\top \ang{\gobs}_{p} -\ln\langle e^{ \params^\top \gobs}\rangle _{\tilde{p}} }.
\label{eq:dual}
\end{align}
Eq.~\eqref{eq:dual} is an unconstrained convex optimization problem over Lagrangian multipliers $\params$ that enforce the $d$ expectation constraints on $\gobs$. %
The objective depends only on the expectations of $\gobs$ and $e^{\params^\top \gobs}$ under the forward and backward processes, which can be easily computed from empirical samples of $\gobs$.
In the special case of antisymmetric observables and a steady-state system without odd variables, Eq.~\eqref{eq:dual} can be estimated using only forward samples since $\ln\langle e^{ \params^\top \gobs}\rangle _{\tilde{p}}=\ln\langle e^{-\params^\top \gobs}\rangle _{p}$. As we discuss in more detail below, a related (but different) variational bound on EP was proposed in Refs.~\cite{kim2020learning,otsubo2022estimating}.

As an example, consider a system with 1000 binary spins measured at two timepoints, with $\gobs$ encoding two-point correlations $x_{i,t}x_{j,t'}$.  
Eq.~\eqref{eq:dual} is an unconstrained convex optimization problem over $1000^2$ variables, which can be solved using standard numerical techniques. Conversely, a naïve estimate of $\ep$ or $\epGvar$ requires inferrence of $>2^{1000}$ probabilities.

So far, we have made no assumptions regarding multipartite structure. Here, we say that the observables are ``multipartite'' if they can be partitioned into blocks, such that only a single block can be active in any given trajectory. In practice, this situation often arises when observables depend on local subsystems and only one subsystem can change state at a given time. We provide a formal definition of multipartite observables in the \emph{End Matter}. There, we show that multipartiteness allows Eq.~\eqref{eq:dual} to be split into smaller optimization problems, each involving a proportional fraction of variables and samples, dramatically reducing the computation and memory required for optimization.

\PRLsection{Maximum likelihood EP decomposition}

As we show in the SM~\cite{SupplementalMaterial}, the distribution $q^*$ that optimizes~\eqref{eq:me0} belongs to an %
exponential family,
\begin{align}
    q_{\params}(\traj) = \tilde p(\traj) e^{{\params}^\top \gobs(\traj)-\ln \langle e^{{\params}^\top \gobs}\rangle_{\tilde{p}}}.\label{eq:expfam}
\end{align}
Specifically, $q^*=q_{\params^*}$ for optimal parameters $\params^*$ that assign maximum likelihood (ML) to forward trajectories:
\begin{align}
    \params^* = \argmax_{\params \in \mathbb{R}^d} \,\big\langle \ln q_{\params}(\traj)\big\rangle_p  = \argmin_{ \params \in \mathbb{R}^d} \,D(p\Vert q_{\params}).
    \label{eq:mlminprob}
\end{align}
This also leads to the information-geometric Pythagorean theorem~\cite{amari2001information}, which decomposes EP into two nonnegative terms:
\begin{align}
   \underbrace{ D(p\Vert \tilde{p}) }_{\ep}=& \underbrace{D(q^* \Vert \tilde p)}_{\epG} +
     \underbrace{D(p\Vert q^*)}_{\epNotG}.
     \label{eq:ep-decomposition}
\end{align}
$\epNotG\vcentcolon= \ep - \epG=D(p\Vert q^*)$ quantifies the EP not captured by $\epG$, expressed via the ML problem $\epNotG=\min_{\params} D(p\Vert q_\params)$.

We may further decompose $\epNotG$ by using that Eq.~\eqref{eq:expfam} implies that $q^*$ obeys
$q^*_{\Trajvar  \vert \Gvar}(\traj\vert \gobs')=\tilde p_{\Trajvar  \vert \Gvar}(\traj\vert \gobs')$, where conditional probabilities are computed using Bayes' rule, $\tilde p_{\Trajvar  \vert \Gvar}(\traj\vert \gobs') = \tilde p(\traj) \delta_{\gobs',\gobs(\traj)}/\tilde p_{\Gvar}(\gobs')$. 
Using the chain rule of KL divergence, we may write $\epNotG$ as a sum of two terms:
\begin{align}
    \epNotG = \underbrace{D(p_{\Gvar}\Vert q^*_{\Gvar})}_{\epGvar - \epG}+\underbrace{D(p_{\Trajvar  \vert \Gvar}\Vert \tilde {p}_{\Trajvar  \vert \Gvar})}_{\ep - \epGvar}\,.
    \label{eq:decomp2}
\end{align}
is the additional EP that could be inferred by constraining  all statistics (not just the mean) of $\gobs$ in Eq.~\eqref{eq:me0}. The second term %
is the EP that cannot be inferred from \emph{any} statistics of $\gobs$.

Our method also supports a hierarchical decomposition of EP, where contributions from interactions of increasing order (e.g., singletons, pairs, triplets) are successively added. This gives a sequence of bounds $0\le \ep_1 \leq \ep_2 \leq \dots \leq\ep$, with $\Sigma_k=D(q^*_k\Vert \tilde p)$ capturing contributions from functions $\gobs(\traj)$ containing interactions up to order $k$. This structure mirrors information-geometric decompositions in equilibrium MaxEnt~\cite{amari2001information,ay2015information} and allows breaking down EP into interpretable contributions, each satisfying $\Sigma_k-\Sigma_{k-1}=D(q^*_k\Vert q^*_{k-1})$.

Finally, our method gives an estimate of trajectory EP, 
\begin{align}
\eptraj(\traj)\vcentcolon= \ln \frac{q_{\params^*}(\traj)}{\tilde p(\traj)}=\params^{*\top}\gobs(\traj)  -\ln\langle e^{ \params^{*\top} \gobs}\rangle _{\tilde{p}}\,. \label{eq:fluctuating-EP-approximation}
\end{align}
This estimator satisfies $\langle \eptraj\rangle_p=\epG$, and it may be shown to provide the best approximation of trajectory EP $\sigma(\traj)$ within the exponential family $q_\params$. Specifically, if we use $\sigma_{\params}(\traj)= \ln [q_{\params}(\traj)/{\tilde p}(\traj)]$ to indicate the approximation given by parameters $\params$ and $\delta\sigma_{\params}(\traj)=\sigma(\traj)-\sigma_{\params}(\traj)$ the corresponding residual, then $\eptraj$ minimizes the expected error~\footnote{Eq.~\eqref{eq:expcost} is derived by using 
$\delta\sigma_{\params}(\traj)= \ln p(\traj)/q_{\params}(\traj)$ and $\langle e^{-\delta\sigma_{\params}}\rangle_p=1$.}:
\begin{align}
D(p\Vert q_{\params}) =\langle \delta\sigma_{\params} + e^{-\delta\sigma_{\params}} - 1 \rangle_p .\label{eq:expcost}
\end{align}
Here, $\delta\sigma_{\params} + e^{-\delta\sigma_{\params}} - 1 \ge 0$ is an information-theoretic loss function that converges to squared error $(\delta\sigma_{\params})^2/2$  as   $\delta\sigma_{\params} \to 0$.

The trajectory EP estimator is exact ($\sigma(\traj) = \eptraj(\traj)$ for all $\traj$) if and only if $\epG=\ep$, since then $p = q^*$. In this case, if the system is in steady state and without odd variables,  we may exploit antisymmetry $\sigma(\traj)=-\sigma(\revtraj)$ to simplify~\eqref{eq:fluctuating-EP-approximation} as
\begin{align}
\sigma(\traj)=\eptraj(\traj)= \params^{*\top}[\gobs(\traj) -\gobs(\revtraj)]/2\,,\label{eq:fluctuating-EP-approximation-as}
\end{align}
which simplifies further to $\params^{*\top}\gobs(\traj)$ when $\gobs$ is antisymmetric.

\PRLsection{Thermodynamic uncertainty relations (TURs)} 
Eq.~\eqref{eq:dual} can be interpreted as a TUR that relates EP and fluctuations of trajectory observables~\cite{horowitz2020thermodynamic}. Consider any trajectory observable that can be expressed as a linear combination $o(\traj)=\params^\top \bm {g}(\traj)$ for some $\params \in \mathbb{R}^d$. The first term in \eqref{eq:dual} is the expectation of $o(\traj)$, while the second term is the cumulant generating function (CGF) of $o(\traj)$ under $\tilde p$. Thus, $\epG$ bounds the fluctuation-discounted expectation of all such $o(\traj)$, and it can be understood as a higher-order TUR~\cite{dechant2020fluctuation} that constrains all cumulants of trajectory observables. This contrasts with quadratic TURs, which constrain only the mean and variance~\cite{li2019quantifying,manikandan2020inferring,dechant2018multidimensional}. 

To relate our approach to quadratic TURs, we expand the CGF as $\ln\langle e^{ \params^\top \gobs}\rangle _{\tilde{p}}  \approx    \params^\top \ang{\gobs}_{\tilde p}  +\params^\top \K\, \params/2$, where $\K$ is the covariance matrix of $\gobs$ under $\tilde p$. Plugging into~\eqref{eq:dual} gives the approximate optimizer $\hat{\params}$ as the solution to the linear system $\K \,\hat{\params}=\ang{\gobs}_{p} -\ang{\gobs}_{\tilde p}\equiv\ang{\gobs}_{p-\tilde p} $, giving  the weaker bound
\begin{align}
    \epG \ge \GepG &\vcentcolon= \ang{\gobs}_{p - \tilde p}^\top \Kinv \ang{\gobs}_{p} - \ln  \ang{e^{\ang{\gobs}_{p-\tilde p}^\top \Kinv  \gobs }  }_{\tilde p}.\label{eq:Gaussian-TUR}
\end{align}
$\GepG$ can be shown to correspond to a single Newton-Raphson step for maximizing \eqref{eq:dual} starting from $\params=\bm 0$~\cite{SupplementalMaterial}. This bound has the advantage of not requiring optimization, only the solution of a linear system. 
The inequality $\epG\ge \GepG$ is tight when $\gobs$ has Gaussian statistics.

We may also compare our results to existing TURs. We consider the special case of steady-state system  without odd variables and antisymmetric $\gobs$. In this case, %
EP may be bounded as (see \emph{End Matter})
\begin{align}
\ep \ge \epGtur \vcentcolon= \ln\left(1+2\langle \gobs\rangle_{p}^{\top}\mathsf{K}_{p}^{-1} \langle \gobs \rangle_{p}\right).
\label{eq:GturDef}
\end{align}
This bound approaches the usual quadratic form, $\epGtur \simeq  2\langle \gobs\rangle_{p}^{\top}\mathsf{K}_{p}^{-1} \langle \gobs \rangle_{p}$ for sufficiently small $\langle \gobs\rangle_{p}^{\top}\mathsf{K}_{p}^{-1} \langle \gobs \rangle_{p}$. 
This regime holds in the short-time limit, since $\langle \gobs \rangle_p$ scales with the observation period.

The bounds~\eqref{eq:Gaussian-TUR}-\eqref{eq:GturDef} may require solving linear systems involving very large covariance matrices. For multipartite observables, it is often possible to split these bounds into sub-problems that involve smaller matrices (see \emph{End Matter}).

\PRLsection{Related work}
In the machine learning literature, Belghazi et al.~\cite{belghazi2018mutual}  proposed an estimator of KL divergence inspired by the ``Donsker-Varadhan'' variational  representation~\cite{donsker1975asymptotic}. When applied to $\ep=D(p\Vert \tilde p)$, it leads to the lower bound
\begin{align}
    \epMine \vcentcolon= \max_{\bm{\phi}\in\mathbb{R}^k} \pr{ \ang{\nnfunc_{\bm{\phi}}}_p -\ln\langle e^{ \nnfunc_{\bm{\phi}}}\rangle _{\tilde{p}} }\,,
    \label{eq:mine}
\end{align}
where $\nnfunc_{\bm{\phi}}(\traj)$ is a nonlinear function parameterized by $\bm{\phi}\in \mathbb{R}^k$ (e.g., output of a neural network) ~\cite{belghazi2018mutual}. 
$\epG$ in Eq.~\eqref{eq:dual} is a special case of $\epMine$ where $\nnfunc_{\bm{\phi}}(\traj)$ contains all linear combinations of $\gobs(\traj)$. The neural-network-based bound~\eqref{eq:mine} has both practical and conceptual differences with respect to Eq.~\eqref{eq:dual}. For instance, it generally involves a difficult non-convex optimization problem, it does not provide a Pythagorean decomposition of EP, and it cannot be interpreted in terms of large deviations. Exploration of $\epMine$ for thermodynamic inference is left as an interesting future direction.

A related variational expression for thermodynamic inference was developed in Kim et al.~\cite{kim2020learning} and Otsubo et al.~\cite{otsubo2022estimating}. In our notation, it is written as 
\begin{align}
\epKim\vcentcolon= \max_{\params}\pr{\params^{\top}\langle \gobs\rangle_{p}-\langle e^{\params^{\top}\gobs}\rangle _{\tilde{p}}+1}\,.\label{eq:fdivdual}
\end{align}
It is possible to consider a neural-network-based version of $\epKim$ by replacing $\params^\top \gobs(\traj)$ with a parameterized function $v_{\bm{\phi}}(\traj)$, as in Eq.~\eqref{eq:mine}, as explored in Refs.~\cite{kim2020learning,otsubo2020estimating}.

Note that $\epG\ge\epKim$, since $-\ln x\ge-x+1$~\cite{belghazi2018mutual,ruderman2012tighter}, therefore $\epG$   always provides a tighter bound on EP than $\epKim$. The two bounds become equivalent in the limit $\langle e^{\params^{\top}\gobs}\rangle _{\tilde{p}}\to 1$, such as the short-time limit $T\to 0$ with antisymmetric observables when $\langle e^{\params^{\top}\gobs}\rangle _{\tilde{p}}=1+O(T)$~\cite{otsubo2022estimating}.  
In general, however, the two bounds give different results, and in the \emph{End Matter}, we provide a simple example where $\epG$  is arbitrarily better than $\epKim$. We note that outside the short-time limit, $\epKim$ does not have a straightforward interpretation in terms of maximum likelihood inference or large-deviations statistics.

Lynn et al.~\cite{lynn2022decomposing,lynn2022emergence,geli2025non} proposed a way to decompose and bound steady-state EP using a different information-theoretic optimization. Although originally focused on local EP in systems with multipartite dynamics, it can be generalized to arbitrary observables $\gobs$ and non-multipartite systems as 
\begin{align}
\epGL & =\min_{q,\tilde q}D(q\Vert\tilde{q})\;\:\text{where}\;\: \langle\gobs\rangle_{q}=\langle\gobs\rangle_{p},\,\tilde q(\traj)=q(\revtraj)\label{eq:lynn}
\end{align}
where the constraint $\tilde{q}(\traj)=q(\revtraj)$ imposes that the trajectory distribution is stationary and without odd variables. 
Like our bound $\epG$, $\epGL$ can be used to generate hierarchical decompositions of EP.  

Although Eq.~\eqref{eq:lynn} resembles Eq.~\eqref{eq:me0}, it differs in that it simultaneously optimizes both arguments of the KL divergence, $q$ and $\tilde{q}$. 
As discussed in the \emph{End Matter}, the optimal distribution in Eq.~\eqref{eq:lynn} is not in an exponential family, and in general it lacks full support. For this reason, Eq.~\eqref{eq:lynn} does not have a tractable dual expression  analogous to our dual~\eqref{eq:dual}, and there is no straightforward way to scale $\epGL$ to large systems. 

Other  work has explored objective~\eqref{eq:lynn} in combination with other types of constraints, e.g., on waiting times~\cite{skinner2021estimating,nitzan2023universal} and hidden Markovian structure~\cite{skinner2021improved,ehrich2021tightest}. 
However, the resulting optimization typically does not scale to high-dimensional or even moderately-sized systems, in part because the considered constraints lead to non-convex optimization problems.

Our Pythagorean theorem~\eqref{eq:ep-decomposition} is related to previous decompositions of EP in interacting systems~\cite{ito2020unified} and systems with nonconservative forces~\cite{ito2024geometric,kolchinsky2024generalized}. 
Moreover, the general form of Eq.~\eqref{eq:me0} is a Maximum Entropy (MaxEnt) problem over trajectory  distributions~\cite{tang2008maximum,marre2009prediction,nguyen2017inverse,cofre2018information}, 
sometimes called ``maximum caliber''~(MaxCal)~\cite{ghosh2020maximum}. 
Related techniques have been used to infer models from which the entropy flow can be estimated, though without ensuring a lower bound on EP~\cite{ishihara2025state}.

Our approach differs in several other ways from earlier work on MaxEnt and MaxCal. First, we minimize the KL divergence relative to an unknown prior distribution $\tilde p$, from which we typically only have samples. 
This makes standard approximation methods for large-scale MaxEnt problems, such as mean-field and Bethe approximations~\cite{welling2003approximate,ricci2012bethe,aguilera2023nonequilibrium}, not directly applicable. 
Second, we care not only about the parameters $\params$, as is typical of MaxEnt ``inverse problems'', but also about bounding EP by the quantity $\epG$. 
Both of these issues are resolved by the dual formulation~\eqref{eq:dual}.

Other notable approaches to EP inference have employed compression algorithms~\cite{roldan2010estimating,roldan2012entropy,ro2022model} and deep learning of probability flows~\cite{boffi2024deep}.

\PRLsection{Examples: Nonequilibrium spin model and Neuropixels dataset}
We illustrate our method on two examples: a nonequilibrium kinetic Ising model~\cite{crisanti1988dynamics,eissfeller1994mean,aguilera2023nonequilibrium} and \emph{in vivo} spike data from the Neuropixels Visual Behavior repository~\cite{allen2022visual}. 
For both examples, we consider a single discrete time step and binary variables, $\traj = (\bm{x}_{0},\bm x_{1})$ where $x_{i,t} \in \{-1, +1\}$ for $i \in \{1,\dots,N\}$. 
We emphasize that our approach also works for continuous systems. In the SM~\cite{SupplementalMaterial}, we illustrate it on a linear Langevin system.

To calculate $\epG$, we sample from steady state using the Monte Carlo method. %
We then optimize Eq.~\eqref{eq:dual} using gradient ascent with Barzilai–Borwein step sizes~\cite{fletcher2005barzilai}. %
We perform early-stopping using held-out validation data, %
which avoids overfitting even in the far-from-equilibrium regime where many reverse transitions are not sampled. Reported EP estimates are computed on held-out test data. 
Details of the data generation, optimization, and analysis can be found in the SM~\cite{SupplementalMaterial} and our code repository~\cite{coderepo}.

The nonequilibrium spin model is specified by transition probabilities 
\begin{align*}
    T(\bm x_{1}\vert \bm x_{0})=\frac{1}{N} \sum_i\Big[ W_i(\bm x_0) \delta_{\bm x_0,  \bm x_{1}^{[i]}} +(1- W_i(\bm x_0) )\delta_{\bm x_0 ,\bm x_{1}} \Big]
\end{align*}
where we introduced the spin-flip operator: $(\bm x^{[i]})_i =-x_i$ and $(\bm x^{[i]})_j =x_j$ for $j \neq i$. The flip probability for spin $i$ is
\begin{align}
W_i(\bm x) &= \dfrac{\exp(-\beta x_{i} \sum_{j:j\neq i}w_{ij} x_{j} )}{2\cosh(\beta \sum_{j:j\neq i} w_{ij} x_{j})}\,,
   \label{eq:spin-flip}
\end{align}
where $\beta$ is an inverse temperature and $w_{ij}$ are (typically asymmetric) coupling parameters. 
We consider the diluted version of the model~\cite{coolen2001statistical,aurell2011message,zhang2012inference} with $k$ average neighbors. Here $w_{ij}=c_{ij}z_{ij}/\sqrt{k}$, with binary connections $c_{ij}\sim \textrm{Bernoulli}(k/(N-1))$ and real-valued weights $z_{ij}\sim \mathcal{N}(0,1)$. 

The steady-state EP in this model  
has a simple closed-form expression, 
$\ep=\beta  \sum_{i \ne j}  w_{ij} \langle{(x_{i,1} -x_{i,0}) x_{j,0}}\rangle_{\rm st}$, 
where $\langle \cdot \rangle_{\rm st}$ indicates expectations under the stationary process $p$. 
This provides ground truth to evaluate our estimators.

\begin{figure}
    \centering
    \begin{tikzpicture}
        \node[anchor=south west] (O) at (0,0)  {};
        \node[anchor=north west] (A) at ([xshift=-0mm]O.north west){\includegraphics[height=4.2cm]{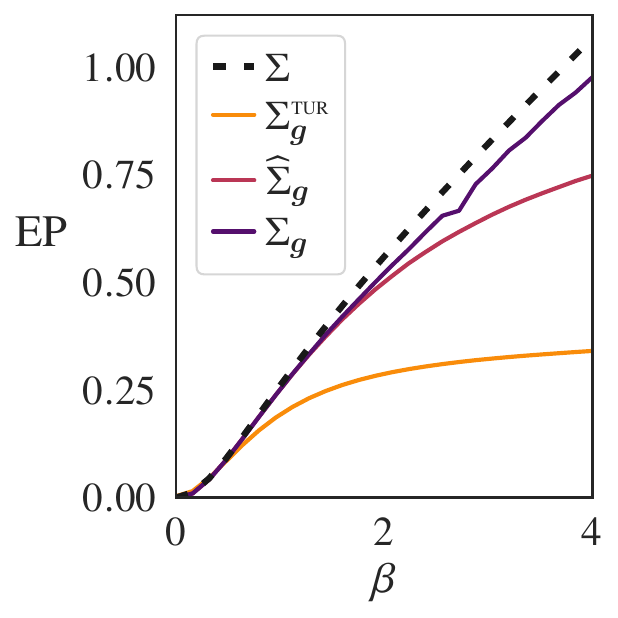}};
        \node[anchor=north west] at ([xshift=-0mm]O.north west) {\textbf{(a)}};
        \node[anchor=north west] (B) at ([xshift=-1mm]A.north east) {\includegraphics[height=4.2cm]{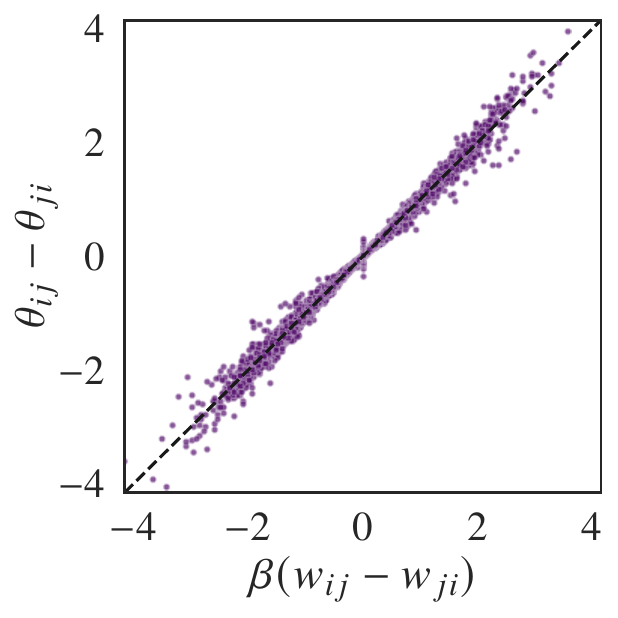}};
        \node[anchor=north west] at ([xshift=1mm]B.north west) {\textbf{(b)}};
    \end{tikzpicture}
    \vspace{-8mm}
    \caption{Disordered nonequilibrium spin model with 1000 spins. \textbf{(a)} Steady-state EP estimates  for different inverse temperatures $\beta$. 
    \textbf{(b)} Asymmetry of inferred parameters,  shown against the true coupling asymmetries in the model for $\beta=2.5$ ($R^2=0.9831$). Estimates are based on $10^9$ state transitions sampled by Monte Carlo.
    }
    \label{fig:spin-model}
\end{figure}

Our observables of interest are time-lagged correlations, 
\begin{align}
    g_{ij}(\traj) =(x_{i,1}-x_{i,0})x_{j,0}\qquad \text{for all}\quad i,j\,,
    \label{eq:obs}
\end{align} 
with conjugate parameters $\theta_{ij}$. %
Because the system has multipartite observables, Eq.~\eqref{eq:dual} can be decomposed into $N$ independent problems, each involving only those transitions where spin $i$ changes, improving computational performance for large systems (see SM). %

Fig.~\ref{fig:spin-model}(a) shows the actual and inferred EP at different $\beta$ for $N=1000$, $k=6$. EP increases with $\beta$, and all estimators ($\epGtur$, $\GepG$, $\epG$) agree in the near-equilibrium regime of small $\beta$. 
Importantly,  $\epG$ provides a tight bound on EP even in the far-from-equilibrium regime of large $\beta$. The gap between $\epG$ and $\GepG$ indicates the onset of highly non-Gaussian statistics from $\beta\approx 2$.

As discussed above, $\epG \approx \ep$ implies that the trajectory EP $\sigma(\traj)$ is closely approximated by $\eptraj$ from Eq.~\eqref{eq:fluctuating-EP-approximation}. 
As shown in the SM, this allows us to use the optimal parameters $\params^*$ to infer the asymmetry of the coupling constants, $\theta_{ij}^*-\theta_{ji}^* \approx \beta(w_{ij}-w_{ji})$. This relation is verified in Fig.~\ref{fig:spin-model}(b).

In the SM, we also consider other coupling matrices for which the model exhibits a nonequilibrium phase transition~\cite{aguilera2023nonequilibrium}. We show that our method accurately infers EP in this regime, a difficult task for existing approximations~\cite{aguilera2021unifying}.

\begin{figure}
    \centering
    \begin{tikzpicture}
        \node[anchor=north west] at (0,0) {\textbf{(a)}};
        \node[anchor=north west] (A) at ([xshift=4mm]0,0) {\includegraphics[width=7cm]{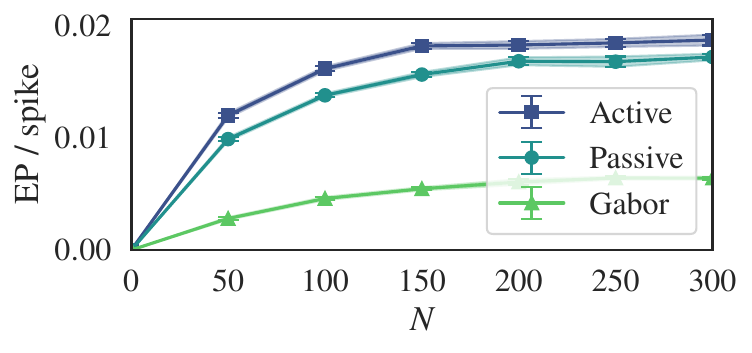}};
        \node[anchor=north west] (B) at ([yshift=0mm,xshift=0mm]A.south west) {\includegraphics[width=7cm]{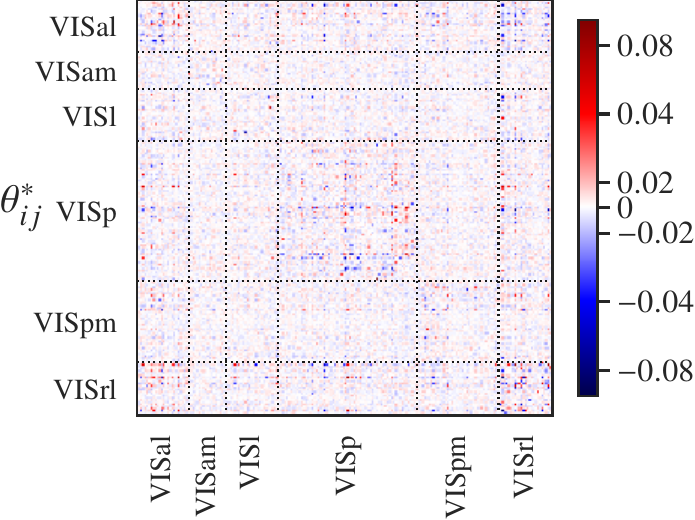}};
        \node[anchor=north west] at ([xshift=-10mm]B.north west) {\textbf{(b)}};
    \end{tikzpicture}
    \vspace{-2mm}
    \caption{\textbf{(a)} Estimated EP per expected number of spikes $R$, in the Neuropixels \emph{Visual Behavior} dataset     for three conditions. Error bars indicate standard error of the mean. \textbf{(b)} A sample of inferred coupling coefficients $\theta^*_{ij}$ grouped by visual area. Here we select 200 neurons with highest firing rate from an active trial. %
    To improve visualization, lower triangle shows $\theta_{ji}^*\equiv-\theta_{ij}^*$ for $i<j$.
}
    \label{fig:neuropixels}
\end{figure}

Next, as an example application to biological data, we estimate EP in the Visual Behavior Neuropixels dataset~\cite{allen2022visual}, which contains spike-train recordings from 81 mice in 103 sessions. In this example, EP is understood as a statistical measure of temporal irreversibility, rather than energetic dissipation~\cite{nartallo2025nonequilibrium}. 
The dataset includes spiking activity from multiple brain regions, including visual cortical areas (VISp, VISl, VISal, VISrl, VISam and VISpm) and subcortical structures.

We analyze data from visual areas during three conditions: active behavior (visual change detection task), passive replay (same stimuli but without task engagement), and Gabor (receptive field characterization with Gabor stimuli and full-field flashes). 
We discretize spike trains into temporal bins of length $10$~ms and verify that most bins contain no more than one spike. In this dataset, 
neurons can update in a parallel (non-multipartite) manner. %

Our observables are defined as %
\begin{align}
    g_{ij}(\traj) =x_{i,1}\,x_{j,0}-x_{i,0}\,x_{j,1}\qquad \text{for all}\quad i<j\,,
    \label{eq:obsneuropixel}
\end{align} 
which naturally represent time-lagged correlations in non-multipartite dynamics. They capture the antisymmetric part of~\eqref{eq:obs} and give more robust numerical results on this dataset.

We estimate EP across different conditions, recording sessions, and system sizes $N$. We randomly select 10 subsets of neurons and estimate $\epG$ for each subset. 
For improved comparison, we normalize EP in each condition by the expected number of spikes per bin, $R = \sum_i \left(1 + \langle x_{i,0} \rangle_{\rm st} \right) / 2$.

Fig.~\ref{fig:neuropixels}(a) shows that EP grows superlinearly with size (note that $R\propto N$). It also shows that the active condition is associated with the largest normalized EP. Fig.~\ref{fig:neuropixels}(b) illustrates the matrix of inferred parameters $\theta_{ij}^*$ for 200 neurons from an active trial, grouped by visual brain area. This matrix specifies a network of functional connectivity~\cite{sporns2013structure}, showing how interactions between individual neurons contribute to temporal irreversibility
of \emph{in vivo} brain dynamics.
This network reveals a clustered organization aligned with anatomical brain regions.

\section*{Author Contributions}
A.K. and M.A. contributed equally to this work.

\section*{Acknowledgements}
\begin{acknowledgments}
M.A. is partly supported by John Templeton Foundation (grant 62828) and Grant PID2023-146869NA-I00 funded by MICIU/AEI/10.13039/501100011033 and cofunded by the European Union, and supported by the Basque Government through the BERC 2022-2025 program and by the Spanish State Research Agency through BCAM Severo Ochoa excellence accreditation CEX2021-01142-S funded by MICIU/AEI/10.13039/501100011033. S.~I. is supported by JSPS KAKENHI Grants  No.~22H01141, No.~23H00467, and No.~24H00834, JST ERATO Grant No.~JPMJER2302, and UTEC-UTokyo FSI Research Grant Program. A.K. is partly supported by John Templeton Foundation (grant 62828) and by the European Union’s Horizon 2020 research and innovation programme under the Marie Sk{\l}odowska-Curie Grant Agreement No.~101068029.
\end{acknowledgments}

\section{Data Availability Statement}

The code used in this study is available in a public repository~\cite{coderepo}. 

\bibliography{references}

\clearpage
\appendix
\setcounter{secnumdepth}{3} 
\section*{End Matter}

\global\long\def\numblocks{k}%
\global\long\def\gobsi{\bm{g}_{i}}%
\global\long\def\zz{\bm{0}}%

\vspace{5pt}

\PRLsection{Multipartite observables} We show that for multipartite observables, our optimization problem can
be split into a set of simpler subproblems. 

We say that observables $\gobs$ are \emph{multipartite} if they can
be decomposed as $\gobs=(\gobs_{1},a_{1},\dots,\gobs_{\numblocks},a_{\numblocks})$
such that: (1) only a single ``block'' of observables $\gobsi$ may be active (non-zero) under any forward or backward
trajectory: $p(\traj)=\tilde{p} (\traj)=0$ whenever $\gobsi(\traj)\ne\zz\wedge \bm{g}_{j}(\traj)\ne\zz$ for some $i\ne j$, and
(2) each $a_{i}(\traj)\vcentcolon= 1-\delta_{\gobsi(\traj),\zz}$ is an indicator variable for  activity of block
$i$.  
For convenience, we use $a_0(\traj)\vcentcolon= 1- \sum_{i=1} a_i(\traj)$ to indicate that  no block is active. (With minor changes, the derivations below
 generalize to the case where  $a_{i}(\traj)=\gamma_{i}(1-\delta_{\gobsi(\traj),\zz})$
for some constants $\gamma_{i}\ne0$.)

Let ${P}_{i}\vcentcolon= \langle a_{i}\rangle_{p}$
be the forward probability that block $i$ is active, and  
 $\condpi(\traj)\vcentcolon= p(\traj\vert\gobs_{i}\ne\bm{0})\vcentcolon= a_{i}(\traj)p(\traj)/{P}_{i}$
the forward trajectory distribution conditioned on block
$i$ being active. $P_{0}=1-\sum_{i}P_{i}$ 
is the probability that no block is active, and $p_0$ is the trajectory distribution conditioned on no block being active.  $\tilde{P}_{i}$, and $\condpirev$
indicate the same quantities under the reverse process $\tilde{p}$. 
In general, $P_{i}$ and $\tilde{P}_{i}$ can be estimated from empirical frequencies, as long as the number of blocks is not very large. 

We now consider our variational expression \eqref{eq:dual} for $\epG$, and we write the optimization variables as $\params=(\bm{\theta}_{1},\dots,\bm{\theta}_{\numblocks},\bm{\lambda})$,
with $\bm{\theta}_{i}$ conjugate to $\gobsi$ and $\lambda_{i}$
conjugate to $a_{i}$ for $i=1...\numblocks$.  For multipartite observables, Eq.~\eqref{eq:dual} 
can be written as
\begin{align}
\epG & =\max_{\params}\sum_{i=1}(\bm{\theta}_{i}^{\top}\langle\gobsi\rangle_{p}+\lambda_i\langle a_{i}\rangle_{p})-\ln Z_{\params,\bm{\lambda}}\nonumber\\
 & =\max_{\params}\sum_{i=1}P_{i}(\bm{\theta}_{i}^{\top}\langle\gobsi\rangle_{p_{i}}+\lambda_{i})-\ln Z_{\params,\bm{\lambda}}.\label{eq:df}
\end{align}
where we introduced $Z_{\params,\bm{\lambda}}\vcentcolon= \langle e^{\bm{\theta}^{\top}\gobs}\rangle_{\tilde{p}}$ for convenience. 
Defining $Z_{\params_{i}}^{i}\vcentcolon= \langle e^{\bm{\theta}_{i}^{\top}\gobsi}\rangle_{\tilde{p}_{i}}$, we may write this term as 
\begin{align}
Z_{\params,\bm{\lambda}}\vcentcolon= \langle e^{\sum_{i=1}\lambda_{i}a_{i}+\bm{\theta}_{i}^{\top}\gobsi}\rangle_{\tilde{p}}=\tilde{P}_{0}+\sum_{i=1}\tilde{P}_{i}e^{\lambda_{i}^{*}}Z_{\params_{i}}^{i}.\label{eq:zzz}
\end{align}
We find the optimal $\bm{\lambda}^{*}$ by taking derivatives
of the objective \eqref{eq:df} with respect to each $\lambda_{i}$:
\begin{equation}
0=P_{i}-\partial_{\lambda_{i}}\ln Z_{\params,\bm{\lambda}^{*}}\implies P_{i}=\tilde{P}_{i}e^{\lambda_{i}^{*}}Z_{\params_{i}}^{i}/Z_{\bm{\theta},\bm{\lambda}^{*}},\label{eq:ddd}
\end{equation}
which gives $\lambda_{i}^{*}=\ln(P_{i}/\tilde{P}_{i})-\ln Z_{\params_{i}}^{i}+\ln Z_{\params,\bm{\lambda}^{*}}$.
Plugging $\tilde{P}_{i}e^{\lambda_{i}^{*}}Z_{\params_{i}}^{i}=Z_{\bm{\theta},\bm{\lambda}^{*}}P_{i}$
into Eq.~\eqref{eq:zzz} and rearranging 
gives $Z_{\params,\bm{\lambda}^{*}}=\tilde{P}_{0}/P_{0}$. Combining
 with Eq.~\eqref{eq:df} and simplifying shows that,
at the optimal $\bm{\lambda}^{*}$, the objective can be written as
\begin{align*}
\epG=D(P\Vert\tilde{P})+\max_{\bm{\theta}_{1},\dots,\bm{\theta}_{\numblocks}}\Big[\sum_{i=1}P_{i}(\bm{\theta}_{i}^{\top}\langle\gobsi\rangle_{p_{i}}-\ln Z_{\params_{i}}^{i})\Big]\,.
\end{align*}
The sum now involves non-overlapping parameter blocks that can be optimized
independently. 

We arrive at our main result for multipartite observables:
\begin{align}
\epG=D(P\Vert\tilde{P})+\sum_{i=1} P_{i}\epGi,\label{eq:lb}
\end{align}
where $\epGi$ is the contribution from block $i=1...\numblocks $, 
\begin{align}
\epGi\vcentcolon= \max_{\bm{\theta}_{i}}\left(\bm{\theta}_{i}^{\top}\langle\gobsi\rangle_{p_i}-\ln\langle e^{\bm{\theta}_{i}^{\top}\gobsi}\rangle_{\tilde{p}_{i}}\right).
\end{align}

Although the decomposition \eqref{eq:lb} increases the number of
optimization problems, each optimization problem is (typically $\sim k$
times) smaller than Eq.~\eqref{eq:dual}, both in terms of the number
of optimization variables and data points needed to estimate
expectations under $\condpi$ and $\condpirev$. In many
cases, we may also bound each $\epGi$ using the optimization-free
bound $\hat{\Sigma}_{\gobs}^{(i)}$ using a smaller covariance matrix
than in Eq.~\eqref{eq:Gaussian-TUR}.

\begin{figure}[t!]
    \centering
    \begin{tikzpicture}
        \node[anchor=south west] (O) at (0,0)  {};
        \node[anchor=north west] (A) at ([xshift=-0mm]O.north west){\includegraphics[height=4cm]{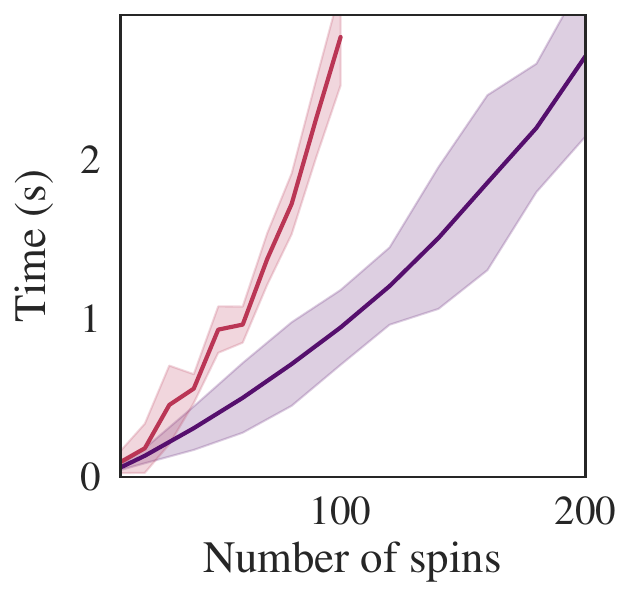}};
        \node[anchor=north west] at ([xshift=-0mm]O.north west) {\textbf{(a)}};
        \node[anchor=north west] (B) at ([xshift=-1mm]A.north east) {\includegraphics[height=4.cm]{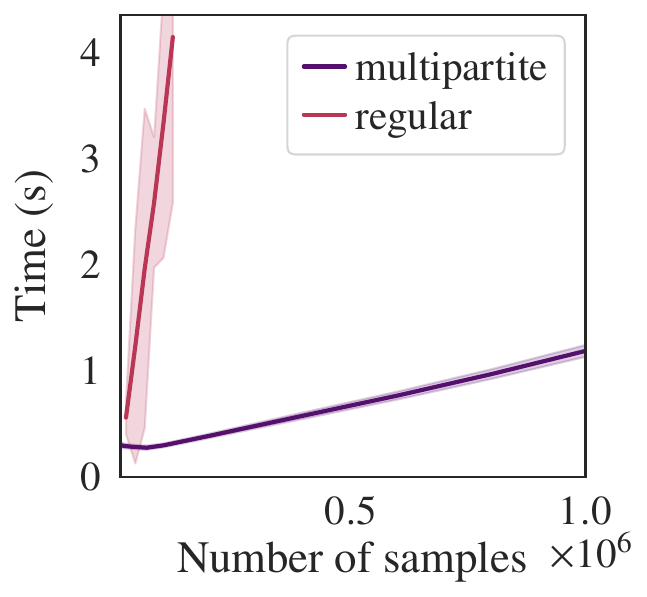}};
        \node[anchor=north west] at ([xshift=-0mm]B.north west) {\textbf{(b)}};
    \end{tikzpicture}
    \vspace{-8mm}
    \caption{{Computation time for $\epG$ using the regular (non-multipartite) optimization~\eqref{eq:dual} versus the multipartite decomposition~\eqref{eq:lb}.}
    \textbf{(a)} Runtime versus number of spins at fixed number of samples $2\times 10^4$.  
    \textbf{(b)} Runtime versus number of samples per spin (system with $40$ spins). 
    Shaded bands: standard deviations over 1000 trials. The end of the ``regular'' curve indicates the point where the GPU runs out of memory.
    Hardware: Intel Core i9-12900KF CPU, NVIDIA GeForce RTX 3050 GPU with 8 GB VRAM.
    }
    \label{fig:computation_time}
\end{figure}

Fig.~\ref{fig:computation_time} demonstrates the improved performance scaling of the multipartite optimization. Here we consider the nonequilibrium spin model across different system sizes (number of spins) and dataset sizes (number of samples). The multipartite method exhibits slower growth in computation time, particularly as the number of samples increases. In contrast, the non-multipartite optimization rapidly exhausts the 8 GB of GPU memory available in our experiments.

We finish by considering the approximation of trajectory EP. For multipartite observables, we can decompose Eq.~\eqref{eq:fluctuating-EP-approximation} as
\begin{align*}
&\eptraj(\traj) = \sum_i  \big[\lambda_i^* a_i(\traj) + \params^{*\top}_i \gobs_i(\traj)\big] - \ln Z_{\params,\bm{\lambda}^*}\\
&= a_0(\traj)\ln \frac{P_0}{\tilde P_0}+ \sum_i  a_i(x) \Big[\ln\frac{P_i}{\tilde P_i} + \params^{*\top}_i \gobs_i(\traj) - \ln Z_{\params^*_i}^i\Big]\end{align*}
where we used expressions of $\lambda_i^*$ and $Z_{\params^*,\bm{\lambda}^*}$ derived above.

When each observable block $\gobsi$ is antisymmetric and the system is in steady state and without odd variables, it can be shown that $P_i = \tilde{P}_i$. This leads to the simplified expressions for estimators of average and trajectory EP:
\begin{align}
\begin{aligned}
\epG&=\sum_{i=1} P_{i}\epGi\\
\eptraj(\traj)&=  \sum_i  a_i(x) \Big[\params^{*\top}_i \gobs_i(\traj) - \ln Z_{\params^*_i}^i\Big]
\end{aligned}\label{eq:lbss}
\end{align}

\vspace{5pt}

\PRLsection{Multidimensional TUR~\eqref{eq:GturDef}}
For a stationary system without odd variables, the ``fluctuation theorem uncertainty relation''~\cite{hasegawa2019fluctuation} states that $e^{\Sigma}-1\ge2\langle\phi\rangle^{2}_p/\mathrm{Var}_p(\phi)$ 
for any antisymmetric scalar observable $\phi(\traj)$. 
Then, for any antisymmetric $\gobs(\traj)$ and any $\bm\alpha \in\mathbb{R}^d$, we may define  $\phi(\traj)\vcentcolon= \bm\alpha^\top\gobs(\traj)$. After some rearranging, this gives
\begin{align}
 \Sigma \ge \ln\big[1+2 (\bm \alpha^\top \langle \gobs \rangle_p )^2 / (\bm \alpha^\top \mathsf K_p \bm \alpha)\big]\,.
\end{align}
Setting $\bm \alpha = \mathsf K_p^{-1} \langle \gobs \rangle_p$ and simplifying gives Eq.~\eqref{eq:GturDef}. 

A similar derivation may be found in Ref.~\cite[Eq.~(8.31)]{peliti2021stochastic}.  A continuous-time version of this bound appeared in Ref.~\cite{dechant2018multidimensional}.

\global\long\def\jU{P_+}%
\global\long\def\jD{P_-}%
\global\long\def\jS{P_{\text{stay}}}%

\begin{figure}
    \centering
    \begin{tikzpicture}
        \node[anchor=south west] (O) at (0,0)  {};
        \node[anchor=north west] (A) at ([xshift=0mm]O.north west){\includegraphics[width=1\columnwidth]{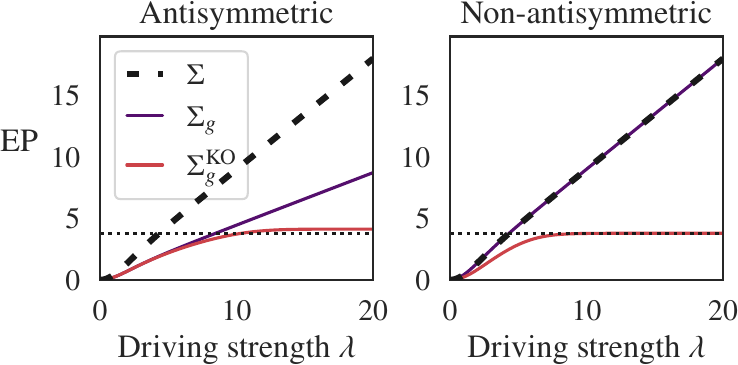}};
        \node[anchor=north west] at ([xshift=3mm]O.north west) {\textbf{(a)}};
        \node[anchor=north west] (B) at ([xshift=4.5cm]A.north west) {\textbf{(b)}};
    \end{tikzpicture}
    \caption{Comparison between our EP bound $\epG$~\eqref{eq:dual} and the variational bound $\epKim$~\eqref{eq:fdivdual} on a 3-state unicyclic system. {(a)} For the non-antisymmetric observable $g(\traj)=1 + \delta_{x_{0}+1,x_{1}}-\delta_{x_{0}-1,x_{1}}$. {(b)} For the antisymmetric observable $g(\traj)=\delta_{x_{0}+1,x_{1}}-\delta_{x_{0}-1,x_{1}} - 1.2 (\delta_{x_{1},1} \delta_{x_{0},0}-\delta_{x_{0},1} \delta_{x_{1},0})$. In both cases, 
    $\epG$ diverges in the irreversible limit $\lambda\to \infty$, while $\epKim$ saturates at a finite value.
    }
    \label{fig:vsneep}
\end{figure}

\vspace{5pt}
\PRLsection{Comparison with $\epKim$ from Refs.~\cite{kim2020learning,otsubo2022estimating}}
Here we compare $\epG$~\eqref{eq:dual} with the variational bound $\epKim$~\eqref{eq:fdivdual} from  Refs.~\cite{kim2020learning,otsubo2022estimating}. 
Using simple
examples, we show that $\epG$ sometimes gives an {arbitrarily} better
bound on EP than $\epKim$. 

We consider a 3-state Markov chain measured at two timepoints $t\in\{0,1\}$, with corresponding states $x_t\in \{0,1,2\}$. The system has uniform unicyclic transition probabilities.  
We parameterize the probability of moving up ($x_{1} = x_0+ 1$  mod 3), down ($x_1 = x_0 - 1$ mod 3), and staying ($x_1=x_0$) as
\begin{align}
\jU= \frac{\kappa}{e^{-\lambda}+1},\quad\jD= \frac{\kappa e^{-\lambda}}{e^{-\lambda}+1},\quad P_{\text{stay}}=1-\kappa \,.
\end{align}.
The parameter $\kappa$ controls dynamical activity while $\lambda$ controls driving strength. The steady state is nonequilibrium if $\lambda \ne 0$.

The system has a uniform steady state with EP $\Sigma = (\jU-\jD)\lambda$. For the estimators, we first consider a single observable $g(\traj) =1 + \delta_{x_{0}+1,x_{1}}-\delta_{x_{0}-1,x_{1}}$, 
whose expectation is $\langle g\rangle _{p}=1+(\jU-\jD)$.
The values of the two estimators %
are
\begin{align}
\epG & =\max_{\theta\in\mathbb{R}}\left[\theta\langle g\rangle _{p}-\ln(\jS e^\theta + \jD e^{2\theta}+\jU )\right]\label{eq:emouropt}\\
\epKim & =\max_{\theta\in\mathbb{R}}\left[\theta\langle g\rangle _{p} - \jS e^\theta - \jD e^{2\theta}-\jU + 1\right]\label{eq:emouropt2}
\end{align}
$\Sigma$, $\epG$ and $\epKim$ all vanish at $\lambda=0$ and increase monotonically in $\lambda > 0$. The optimization problem~\eqref{eq:emouropt} can be solved in closed form to find $\epG=\ep$ with optimal parameter $\theta^* = \ln {\jU}/{\jD} = \lambda $, thus $\ep=\epG \to\infty$ in the irreversible limit $\lambda\to\infty$. The expression for $\epKim$ is more complicated, but it can be shown that %
it saturates at the finite value $\epKim\to (1+\kappa)2\tanh^{-1}\kappa  - 2\kappa$ in the irreversible limit. 
Numerical results are shown in Figure~\ref{fig:vsneep}(a) for $\kappa=9/10$ and a range of driving strengths. 
It is seen that $\epKim$ remains finite  while  $\epG$ 
diverges for large $\lambda$. 

The above example features a non-antisymmetric observable, but the effect also holds for some antisymmetric observables. As an example, we may consider the observable $g(\traj)=\delta_{x_{0}+1,x_{1}}-\delta_{x_{0}-1,x_{1}} - 1.2 (\delta_{x_{1},1} \delta_{x_{0},0}-\delta_{x_{0},1} \delta_{x_{1},0})$. In this case, $\epG < \ep$, but our estimator still captures a diverging amount of EP in irreversible limit,  $\epG\to \infty$  as $\lambda \to \infty$. On the other hand, $\epKim$ saturates at the finite value $\approx 4.13$ as $\lambda \to \infty$ ($\kappa=9/10$).  Figure~\ref{fig:vsneep}(b) displays the numerical values of the two estimators for this antisymmetric observable.

\vspace{5pt}
\PRLsection{Comparison with Lynn et al.~\cite{lynn2022emergence}}
We consider the optimization problem that defines $\epGL$~\eqref{eq:lynn}, our generalization of the estimator from Ref.~\cite{lynn2022emergence}. %
The partial derivative of the objective with respect to $q(\traj)$ is  
\begin{align*}
    \partial_{ q(\traj)} D(q\Vert \tilde q)=1 + \ln \frac{q(\traj) }{\tilde q(\traj) } - \frac{\tilde q(\traj) }{q(\traj) }\,.
\end{align*}
where we used the condition $\tilde q(\traj) = q(\revtraj)$. 
This expression does not diverge even when a pair of probabilities $q(\traj)$ and $q(\revtraj)$ approach zero. 
Therefore, in general, the optimal distribution in Eq.~\eqref{eq:lynn} may not have full support, instead lying on the boundary of the set of probability distributions. 
Using some numerical examples, we have verified that the optimization~\eqref{eq:lynn} often returns solutions without full support.

For comparison, the partial derivative of our MaxEnt objective~\eqref{eq:me0}, $\partial_{ q(\traj)} D(q\Vert \tilde p)=1 + \ln [{q(\traj) }/{\tilde p(\traj) }]$ diverges to $-\infty$ as $q(\traj)\to 0$ for any $\traj$ where $\tilde p(\traj)> 0$. This implies that a point on the boundary cannot be the minimizer, thus the optimal solution will lie in the relative interior of the feasible set.

This difference has significant consequences for the tractability of the two optimization problems.
The fact that strict positivity is enforced by our objective~\eqref{eq:me0} allows us to restate the optimization in terms of a tractable dual problem~\eqref{eq:dual}. %
On the other hand, the dual formulation of~\eqref{eq:lynn} requires an exponential number of nonnegative Lagrange multipliers, one for each pair of trajectories $\traj$ and $\revtraj$ to guarantee nonnegativity of $q(\traj)$ and $q(\revtraj)$. Therefore, the dual formulation of \eqref{eq:lynn} involves an intractable constrained optimization problem over an exponential number of parameters.

\nocite{boyd2004convex, cover1999elements, barzilai1988two,fletcher2005barzilai, lin1999newton, aguilera2021unifying, aguilera2023nonequilibrium, allen2022visual}

\end{document}


\title{SUPPLEMENTAL MATERIAL}

\maketitle

Here we provide more details of our derivations, methods, and the two examples considered in the paper: the nonequilibrium spin model and the Neuropixels spike dataset.

\section{Derivation of dual variational principle}
\label{app:dual}

We recall our nonequilibrium MaxEnt problem
\begin{align}
\epG & := \min_{q}D(q\Vert\tilde{p})\quad\text{subject to} \quad\ang{\gobs}_{q}=\ang{\gobs}_{p}\,.
\label{eq:me0}
\end{align}
It has the associated Lagrangian 
\begin{align*}
\mathcal{L}(q,\params,\lambda) = D(q\Vert \tilde{p}) 
+\params^\top (\ang{\gobs}_p-\ang{ \gobs}_q)
     + \lambda (1 - \langle 1\rangle_{q})\,,
\end{align*}
where $\params\in\mathbb{R}^d$ enforces the expectation constraints and $\lambda\in\mathbb{R}$ enforces the normalization constraint $\langle 1\rangle_q = \sum_{\traj} q(\traj)=1$. We do not have to include Lagrange multipliers for the nonnegativity constraints $q(\traj)\ge 0$ because,  as we show below, the optimal distribution will always be in an exponential family and have the same support as $\tilde{p}$.

To find the critical points of the Lagrangian, we evaluate the gradient of $\mathcal{L}$ with respect to $q(\traj)$,%
\begin{equation}
    1 + \ln q^*(\traj)  - \ln \tilde p(\traj) -  \params^\top \gobs(\traj) - \lambda = 0,
    \label{eq:lagcrit}
\end{equation}
giving the solution 
$q^*(\traj)
= \tilde  p(\traj)e^{\params^\top  \gobs(\traj) + \lambda - 1} $.  
The multiplier $\lambda$ is determined by the  normalization condition $\langle 1\rangle_{q^*}=1$ and can be eliminated, yielding the exponential-family form 
\begin{align}
    q^*(\traj)=q_\params(\traj) = \tilde  p(\traj) e^{\params^\top  \gobs(\traj) - \ln\langle e^{\params^\top  \gobs}\rangle_{\tilde p}}.
     \label{eq:exponential-distribution}
\end{align}

The constrained optimization~\eqref{eq:me0} can be written in terms of the Lagrangian as
\begin{align}
\epG = \min_{q} \max_{\params, \lambda} \mathcal{L}(q,\params,\lambda).
\end{align}
Because the objective in \eqref{eq:me0} is convex and $p$ is a feasible point on the relative interior of the feasible set, we may use convex duality~\cite{boyd2004convex} to write this optimization as
\begin{align}
    \epG = \max_{\params, \lambda} \min_{q}      \mathcal{L}(q,\params,\lambda).
    \label{eq:appdual}
\end{align}
The inner optimization can be solved to give~\eqref{eq:exponential-distribution}, which can be plugged back into \eqref{eq:appdual} and simplified to give  our dual form,
\begin{align}
\epG  = \max_{\params \in \mathbb{R}^d} \pr{ \params^\top \ang{\gobs}_{p} -\ln\langle e^{ \params^\top \gobs}\rangle _{\tilde{p}} }.
\label{eq:dual}
\end{align}
The exponential-family distribution $q_{\params^*}$ for the optimal parameters $\params^*$ is also the optimizer of the MaxEnt problem~\eqref{eq:me0}.

The optimal $\params^*$ in Eq.~\eqref{eq:dual} may also be derived by maximizing expected likelihood of forward trajectories in the exponential family~\eqref{eq:exponential-distribution},
\begin{align}
\params^* = \argmax_{\params} \,\big\langle \ln q_{\params}(\traj)\big\rangle_p
\label{eq:appml}
\end{align}
To see why, observe that the objective of the maximum likelihood (ML) problem is concave and that the optimal $\params^*$ satisfies
\begin{align}
\bm 0 = \nabla_\params {\big\langle \ln q_{\params}(\traj)\big\rangle_p} \big\vert_{\params=\params^*} = \ang{\gobs}_p - \ang{\gobs}_{q_{\params^*}}\,.
\end{align}
This implies that $q_{\params^*}$ satisfies the expectation constraints in \eqref{eq:me0} while also being at a critical point of the Lagrangian \eqref{eq:lagcrit}. Therefore, it must be the optimizer of the MaxEnt problem~\eqref{eq:me0}. The maximum likelihood can also be put in the form of an information-projection, 
\begin{align}
\params^* = \argmin_{\params} \,D(p\Vert q_{\params}),\label{eq:iproj}
\end{align}
since the objectives in Eqs.~\eqref{eq:appml} and \eqref{eq:iproj} are equal (up to sign change and additive constant).

\section{Coarse-grained KL divergence bounds}

For clarity, in this section we explicitly indicate trajectory distributions like $p,\tilde p,q,\dots$ with subscripts like $p_{\Trajvar},\tilde p_{\Trajvar}, q_{\Trajvar}, \dots$. 
We denote by $\gobs(\traj)$ the vector of $d$ observables associated with each trajectory $\traj$. Given distribution $p_{\Trajvar}$, we use the notation $p_{\Trajvar \Gvar}(\traj,\gobs^\prime)=p_{\Trajvar}(\traj)\delta_{\gobs^\prime, \gobs(\traj)}$ to indicate the associated joint distribution over trajectories and observables (and similarly for other distributions such as $\tilde p_{\Trajvar\Gvar},q_{\Trajvar\Gvar},\dots$). Such joint distributions induce marginal distributions over observables, $p_{\Gvar}(\gobs^\prime)=\sum_{\traj} p_{\Trajvar}(\traj) \delta_{\gobs^\prime, \gobs(\traj)}$, as well as corresponding conditional distributions $p_{\Trajvar \vert \Gvar}$ and $p_{\Gvar\vert \Trajvar}$, computed according to the usual rules of probability calculus. %
For continuous-state systems, sums should be replaced by integrals and the Kronecker delta by the Dirac delta.

Let us consider the optimization problem that defines $\epG$,
\begin{align}
    \epG := \min_{q} D(q_{\Trajvar} \Vert \tilde{p}_{\Trajvar}) \quad \text{subject to} \quad \langle \gobs  \rangle_{q_{\Trajvar}} = \langle \gobs  \rangle_{p_{\Trajvar}}\,.
    \label{eq:variational-constraint}
\end{align}
We introduce the trajectory distribution %
$q^\prime_{\Trajvar}(\traj):=p_{\Gvar}(\gobs(\traj))\tilde p_{\Trajvar \vert \Gvar}(\traj \vert \gobs(\traj))$. 
It is easy to verify by marginalization that $q^\prime_{\Trajvar}$ satisfies the constraint $\langle \gobs\rangle_{q^\prime_{\Trajvar}}=\langle \gobs\rangle_{p_{\Trajvar}}$, thus %
$\epG\le D(q^\prime_{\Trajvar} \Vert \tilde p_{\Trajvar})$. We also have the identities
\begin{align*}
     D(q^\prime_{\Trajvar}\Vert \tilde{p}_{\Trajvar})  &=      D(q^\prime_{\Trajvar}\Vert \tilde{p}_{\Trajvar})+D( q^\prime_{\Gvar\vert \Trajvar} \Vert \tilde p_{\Gvar\vert \Trajvar})\\
     &= D(q^\prime_{\Trajvar \Gvar}\Vert \tilde{p}_{\Trajvar \Gvar}) =D(p_{\Gvar}\tilde p_{\Trajvar \vert \Gvar}\Vert \tilde{p}_{\Gvar}\tilde{p}_{\Trajvar\vert \Gvar})\\
     &=D(p_{\Gvar} \Vert \tilde p_{\Gvar}) + D(\tilde p_{\Trajvar \vert \Gvar} \Vert \tilde p_{\Trajvar \vert \Gvar})=D(p_{\Gvar} \Vert \tilde p_{\Gvar})\,.
\end{align*}
In the first line, we used $D(q^\prime_{\Gvar\vert \Trajvar} \Vert \tilde p_{\Gvar\vert \Trajvar})=0$ since $q^\prime_{\Gvar\vert \Trajvar}(\gobs^\prime \vert \traj)=\tilde{p}_{\Gvar\vert \Trajvar}(\gobs^\prime \vert \traj)=\delta_{\gobs^\prime,\gobs(\traj)}$; in the second line, we used the chain rule of KL divergence and the definition of $q^\prime_{\Trajvar}$; in the last line, we again used the chain rule of KL divergence. 
Furthermore, by the data processing inequality~(DPI)~\cite{cover1999elements}, the KL divergence  obeys the bound $\ep = D(p_{\Trajvar} \Vert \tilde{p}_{\Trajvar}) \ge D(p_{\Gvar}\Vert \tilde{p}_{\Gvar})$. 
This follows from the fact that the observable map $\gobs(\traj)$ is a (potentially many-to-one) function of the trajectory $\traj$, which constitutes a form of coarse-graining. %
Combining these results, we obtain the following chain of inequalities: 
\begin{align}
    \epG
    \le \underbrace{D(p_{\Gvar} \Vert \tilde{p}_{\Gvar})}_{ \epGvar} \le \underbrace{D(p_{\Trajvar}\Vert \tilde{p}_{\Trajvar})}_{\ep}\,.
\end{align}

\section{Optimization details}
\label{app:optimization}
\subsection{Gradient ascent}

We used gradient ascent to optimize our objective $L(\params)= \params^\top \ang{\gobs}_p -\ln\langle e^{ \params^\top \gobs}\rangle _{\tilde{p}}$ from Eq.~\eqref{eq:dual} and generate the results for the nonequilibrium spin model and the Neuropixels dataset. 
In gradient ascent, the parameters are updated iteratively as 
\begin{align}
\params_{k+1} = \params_k + \alpha_k \nabla L(\params_k),
\label{eq:gradascent-sm}
\end{align}
where $\alpha_k$ is the step size. 
The gradient of the objective  %
is
\begin{align}
    \nabla L(\params)=\ang{\gobs}_p - \ang{\gobs}_{q_{\params}}\,,
    \label{eq:gradsm}
\end{align}
where $\ang{\gobs}_{q_{\params}}$ is the expectation of $\gobs$ under distribution $q_\params(\traj) = \tilde{p} (\traj) e^{ \params^\top \gobs(\traj) - \ln \langle e^{ \params^\top \gobs} \rangle_{\tilde{p}} }$. It may also be written in terms of expectations under $\tilde p$ as
\begin{align}
\langle  \gobs \rangle_{q_\params} & =\frac{\langle e^{\params^\top \gobs(\traj)}\gobs(\traj)\rangle_{\tilde p}}{\langle e^{\params^\top \gobs(\traj)}\rangle_{\tilde p} }. \label{eq-sm:gradient}
\end{align}

For gradient ascent optimization, we used $\params_0 = \bm{0}$ and $\alpha_0 = 1/N$ as the starting parameters ($N$ is the dimensionality of the system).  %
We chose the step sizes according to the ``short step'' Barzilai–Borwein rule~\cite{barzilai1988two,fletcher2005barzilai}, 
\begin{align}
\alpha_k = \frac{(\params_k - \params_{k-1})^\top (\nabla L(\params_k) - \nabla L(\params_{k-1}))}{\Vert \nabla L(\params_k) - \nabla L(\params_{k-1})\Vert^2}.
\label{eq:sm-BB}
\end{align}
This rule estimates curvature information using previous iterations, without requiring explicit Hessian computations.  %
It %
often leads to fast convergence %
on convex optimization problems. 

To prevent overfitting, we employed a standard holdout strategy. The data was randomly split into $70\%$ for training, $20\%$ for validation, and $10\%$ for testing. The training set was used to estimate the gradients of the objective. %
Optimization was halted %
once the objective on the validation set failed to improve for 10 consecutive iterations (early stopping). In addition, to prevent overfitting, optimization was also halted when estimated EP in training or validation was larger than the logarithm of the number of samples, $\ln n_\mathrm{samples}$. This cutoff is motivated by the fact that it is not possible to reliably estimate KL divergence values larger than $\ln n_\mathrm{samples}$, %
since events with probability less than $1/n_\mathrm{samples}$ are unlikely to be observed. 
Finally, we selected the parameters $\params^*$ that achieved the highest validation objective during training. We used these parameters to evaluate the objective on a held-out test set, which we reported as the final estimate of entropy production.
This procedure gave reliable results and prevented overfitting, even in the far-from-equilibrium regime where some reverse transitions were too rare to be observed given realistic sample sizes. 

\subsection{Newton-Raphson optimization}

The bound $\GepG$ from Eq.~\eqref{eq:Gaussian-TUR} can be interpreted in terms of Newton-Raphson optimization. To see this, note that the Newton-Raphson update for maximizing an objective function $L(\params)$  is 
\begin{align}
    \params_{k+1}  =\params_k-\left( \nabla^2 L(\params_k) \right)^{-1} \nabla L(\params_k).
    \label{eq:newtonregular}
\end{align}
For the objective  $L(\params)= \params^\top \ang{\gobs}_p -\ln\langle e^{ \params^\top \gobs}\rangle _{\tilde{p}}$, the gradient is given by Eq.~\eqref{eq:gradsm}. The Hessian  $\nabla^2 L(\params) =  -\mathsf{K}_{q_{\params}}$ is equal to 
the covariance of $\gobs$ under distribution $q_\params(\traj) = \tilde{p} (\traj) e^{ \params^\top \gobs(\traj) - \ln \langle e^{ \params^\top \gobs} \rangle_{\tilde{p}} }$, 
\begin{align}
(\mathsf{K}_{q_{\params}})_{nm}=\langle{g_n  g_m}\rangle_{q_\params}-\langle  g_n \rangle_{q_\params}\langle  g_m \rangle_{q_\params}=\frac{\langle e^{\params^\top \gobs} g_n g_m\rangle_{\tilde p}}{\langle e^{\params^\top \gobs}\rangle_{\tilde p}}-\langle  g_n \rangle_{q_\params}\langle  g_m \rangle_{q_\params}.
\label{eq-sm:hessian}
\end{align}
Taking a single step from $\params_k = \bm 0$ recovers  
$\GepG$~\eqref{eq:Gaussian-TUR}, because $\params_{k+1} = \mathsf{K}_{\tilde{p}}^{-1}(\ang{\gobs}_{p}-\ang{\gobs}_{\tilde{p}} ) =\hat{\params}$.

\section{NONEQUILIBRIUM SPIN MODEL}

\subsection{Model and data generation}
To define our nonequilibrium spin model, 
we randomly sample (and quench)  the coupling parameters $w_{ij}$ as a product of two independently sampled random variables: $w_{ij} = c_{ij} z_{ij} / \sqrt{k}$. Here, $z_{ij} \sim \mathcal{N}(0, 1)$ are Gaussian-distributed weights, while $c_{ij} \sim \mathrm{Bernoulli}(k/(N-1))$ are binary variables that enforce sparsity by randomly selecting a subset of active connections. This construction yields a sparse network with quenched disorder. We use an average in-degree of $k = 6$.

We use Monte Carlo to draw $10^6 \times N$ samples from $10^3$ independent restarts. Each restart begins on a random initial configuration and then undergoes a burn-in period of $10^5 \times N$ Glauber updates according to Eq.~\eqref{eq:spin-flip}. For computational efficiency, for each visited configuration, we independently sample a Glauber update for each of the $N$ spins before moving on to the next configuration. To reduce temporal correlations, we discard $N$ Glauber steps between sampling intervals.

\subsection{Optimization using multipartite assumption}
\label{app-multipartite-optimization}

The nonequilibrium spin model satisfies the assumptions of our multipartite analysis, as discussed in the \emph{End Matter}. Specifically,  
the observables~\eqref{eq:obs} are multipartite because $g_{ij}(\traj) \ne 0$ only if spin $i$ changes state in transition $\traj$, and only a single spin can change state at any one time (i.e., the dynamics are multipartite). Therefore, the observables $\gobs(\traj)$  
are grouped into $N$ blocks,
\begin{align}
\gobs_{i}=(g_{i,1},\dots,g_{i,i-1},g_{i,i+1},\dots,g_{i,N})\,.
\label{eq:smGblocks}
\end{align}
Each block $i$ contains $(N-1)$ observables ($g_{ij}$ for all $j$ such that $j \ne i$).
The diagonal observables $g_{ii}$ are proportional to the indicator function for flips of spin $i$
$$g_{ii}(\traj)=(x_{i,1}-x_{i,0})x_{i,0}=-2(1-\delta_{x_{i,0},x_{i,1}})=-2 a_i(\traj),$$
as required by our analysis.  
Thus, we may use the method discussed in the \emph{End Matter} to split our optimization problem into $N$ independent subproblems,  $\epG =\sum_{i=1} P_i \epGi$. Here we used that
the observables $\gobs_{i}$ are antisymmetric and the system is in steady state and without odd variables, thus $D(P\Vert \tilde P)=0$.

Each subproblem involves an optimization over $N-1$ parameters. We solve it using gradient ascent with Barzilai–Borwein steps. 
For moderately sized $N$, in principle, the objective can be optimized using second-order methods such as Newton's method. In practice, we found that trust-region Newton optimization~\cite{lin1999newton} gave  similar results to gradient ascent.

We also used the multipartite assumption to simplify the computation of the multivariate TUR bound $\epGtur$. 
Using the chain rule of KL divergence, EP can be written as 
\begin{align*}
\ep := D(p\Vert\tilde{p})=D(P\Vert\tilde{P})+\sum_{i=0} P_{i}\Sigma^{(i)}\,,
\end{align*}
where $\ep^{(i)}=D(p_{i}\Vert\tilde{p}_{i})$ and  $D(P\Vert\tilde{P})=\sum_{i=0}^{N} P_{i}\ln(P_{i}/\tilde{P}_{i})$
is the KL divergence between block-activity under forward and reverse
 processes.  Since $\gobs_i$ are antisymmetric and the system is in steady state, we have $P_i=\tilde P_i$. In addition, $\ep^{(0)}=D(p_0\Vert \tilde p_0)=0$ since the state of the spin system does not change if no spin flips. Thus, we may simply write $\ep=\sum_{i=1} P_{i}\ep^{(i)}$. We bound each term $\ep^{(i)}$ %
using the multivariate TUR expression~\eqref{eq:GturDef}.

\subsection{Steady-state EP}

The steady-state EP is computed using the steady-state distribution, which satisfies $\pst(\bm x_{1})=\sum_{\bm x_0}T(\bm x_{1}\vert\bm x_{0})\pst(\bm x_0)$.  In steady state, the forward and reverse trajectory distributions are 
\begin{align*}
p(\traj) = T(\bm x_{1}\vert\bm x_{0})\pst(\bm x_0),\qquad \tilde{p}(\traj) = T(\bm x_{0}\vert\bm x_{1})\pst(\bm x_1).
\end{align*}
Using these distributions, we may compute the EP as
\begin{align}
    \ep=D(p\Vert \tilde p) &=  \sum_{\bm x_0 ,\bm x_1} \pst (\bm x_{0}) T(\bm x_{1}\vert\bm x_{0}) \ln \frac{\pst (\bm x_{0}) T(\bm x_{1}\vert\bm x_{0})}{\pst (\bm x_{1}) T(\bm x_{0}\vert\bm x_{1})}\nonumber\\
     &=  \sum_{\bm x_0 ,\bm x_1} \pst (\bm x_{0}) T(\bm x_{1}\vert\bm x_{0}) \ln \frac{  T(\bm x_{1}\vert\bm x_{0})}{  T(\bm x_{0}\vert \bm x_{1})}\nonumber\\
&= \sum_{\bm x_0}\sum_i\pst (\bm x_{0}) T(\bm x_{0}^{[i]}\vert \bm x_{0}) \ln \frac{  T(\bm x_{0}^{[i]}\vert \bm x_{0})}{ T(\bm x_{0}\vert \bm x_{0}^{[i]})}.
\end{align}
In the second line, we used that the change of Shannon entropy vanishes in steady state, $$\sum_{\bm x_0, \bm x_1} \pst (\bm x_{0}) T(\bm x_{1}\vert \bm x_{0}) \ln \frac{\pst (\bm x_{0}) }{\pst (\bm x_{1})} 
= \sum_{\bm x_0}  \pst (\bm x_{0}) \ln \pst (\bm x_{0}) - \sum_{\bm x_1}  \pst (\bm x_{1}) \ln \pst (\bm x_{1})=0.$$ 
In the third line, we used the multipartite structure of the dynamics. 

Using Eq.~\eqref{eq:spin-flip} in the main text, the ratio of transition probabilities for trajectory $(\bm x_0, \bm x_0^{[i]})$ may be written as
\begin{align}
 \ln \frac{ T(\bm x_{0}^{[i]}\vert \bm x_{0})}{T(\bm x_{0}\vert \bm x_{0}^{[i]})} = \ln \frac{W_i (\bm x_0)}{W_i (\bm x^{[i]}_0)}= \beta \sum_{j:j\ne i} w_{ij}(x_{i,0}^{[i]} x_{j,0} -x_{i,0} x_{j,0}),
\label{eq:ratioT}
\end{align}
where we use $x^{[i]}_{i,0} = -x_{i,0}$ 
and cancel the $\cosh (\beta \sum_{j:j\ne i} w_{ij} x_{j,0})$ terms. %
Then, we may write the steady-state EP as 
\begin{align}
    \ep&= \sum_{i\ne j} \sum_{\bm x_0}  \pst (\bm x_{0}) T(\bm x_{0}^{[i]}\vert \bm x_{0})
    \beta w_{ij}(x_{i,0}^{[i]} x_{j,0} -x_{i,0} x_{j,0}) \nonumber\\
    &= \sum_{i \ne j} \sum_{\bm x_0,\bm x_1}   \pst (\bm x_{0}) T(\bm x_{1}\vert \bm x_{0})
    \beta w_{ij}(x_{i,1} x_{j,0} -x_{i,0} x_{j,0}) \nonumber,
\end{align}
where we used that
$x_{i,1} x_{j,0} -x_{i,0} x_{j,0}=0$ for $\bm{x}_{1} \neq \bm x_0^{[i]}$. Finally, introducing the notation $\langle \cdots \rangle_{\rm st}=\sum_{\bm x_0,\bm x_1} \cdots \pst(\bm x_0)T(\bm x_1\vert \bm x_0)$, we may write the steady-state EP as
\begin{align}
    \ep&=\beta  \sum_{i\ne j}  w_{ij} \ang{(x_{i,1} -x_{i,0}) x_{j,0}}_{\rm st}.
    \label{eq:steady-stateEP}
\end{align}
In terms of the observables $\gobs$ from Eq.~\eqref{eq:obs},  defined as  \begin{align}
    g_{ij}(\traj) =(x_{i,1}-x_{i,0})x_{j,0}\,,
    \label{eq:smobs}
\end{align}
the steady-state EP may be expressed as 
$\ep =\beta  \sum_{i \ne j}  w_{ij} \ang{g_{ij}}_{\rm st}$.

\subsection{Relationship between coupling coefficients $w_{ij}$ and parameters $\theta_{ij}^{*}$}

We now derive the relationship between the coupling coefficients
$w_{ij}$ and the optimal parameters $\theta_{ij}^*$. 
We use the following expression for the trajectory EP of transition
$\bm{x}_0\to\bm{x}_0^{[i]}$,
\begin{align}
\sigma(\bm{x}_0,\bm{x}_0^{[i]})&=\ln\frac{T(\bm{x}_0^{[i]}\vert\bm{x}_0)\pst(\bm{x}_0)}{T(\bm{x}_0\vert\bm{x}_0^{[i]})\pst(\bm{x}_0^{[i]})}=-2\beta\sum_{j:j\ne i}w_{ij}x_{i,0}x_{j,0}+\ln\frac{\pst(\bm{x}_0)}{\pst(\bm{x}_0^{[i]})},
\label{eq:appzz0}
\end{align}
as follows from Eq.~\eqref{eq:ratioT} and
$x_{i,0}^{[i]}=-x_{i,0}$. 

We consider the sum of $\sigma$ over a cyclic sequence
of states $\bm{x}_0\to\bm{x}_0^{[i]}\to\bm{x}_0^{[i,k]}\to\bm{x}_0^{[k]}\to\bm{x}_0$, where $\bm{x}_0^{[i,k]}$ indicates state $\bm{x}_0$ with spins $i$ and $k$ flipped. For any initial $\bm x_0$ and spins $i\ne k$, this gives
\begin{align}
&\qquad\qquad\quad\sigma(\bm{x}_0,\bm{x}_0^{[i]})\;\;+\;\;\sigma(\bm{x}_0^{[i]},\bm{x}_0^{[i,k]})\;\;+\;\;\;\sigma(\bm{x}_0^{[i,k]},\bm{x}_0^{[k]})\;\;\;\;+\;\;\;\;\sigma(\bm{x}_0^{[k]},\bm{x}_0)\label{eq:smsumAA}\\
& =-2\beta\left[\sum_{j: j \ne i} w_{ij} x_{i,0} x_{j,0} + \sum_{j: j \ne k} w_{kj} x_{k,0}^{[i]} x_{j,0}^{[i]} + \sum_{j: j \ne i} w_{ij} x_{i,0}^{[i,k]} x_{j,0}^{[i,k]} + \sum_{j: j \ne k} w_{kj} x_{k,0}^{[k]} x_{j,0}^{[k]} \right]\nonumber\\
&= -2\beta\left[\sum_{j: j \ne i} w_{ij} x_{i,0} x_{j,0} + \sum_{j: j \ne k} w_{kj} x_{k,0} x_{j,0}^{[i]} -\sum_{j: j \ne i} w_{ij} x_{i,0} x_{j,0}^{[i,k]} - \sum_{j: j \ne k} w_{kj} x_{k,0} x_{j,0}^{[k]}\right]\nonumber\\
&= -2\beta\left[\sum_{j: j \ne i} w_{ij} x_{i,0} x_{j,0} + \sum_{j: j \ne k} w_{kj} x_{k,0} x_{j,0}^{[i]} -\sum_{j: j \ne i} w_{ij} x_{i,0} x_{j,0}^{[k]} - \sum_{j: j \ne k} w_{kj} x_{k,0} x_{j,0}\right]\nonumber\\
&=- 4\beta(w_{ik}-w_{ki})x_{i,0}x_{k,0}\,.\label{eq:appSsum0}
\end{align}
{The sum of the terms $\ln[\pst(\bm{x}_0)/\pst(\bm{x}_0^{\prime})]$
 cancels. }

As shown in Fig.~\ref{fig:spin-model}(a), our lower bound captures nearly all of
EP, $\ep\approx\epG$. Therefore, we may use Eq.~\eqref{eq:fluctuating-EP-approximation-as} to estimate the trajectory EP for transition $\bm{x}_0 \to \bm{x}_0^{[i]}$ in terms of the optimal parameters $\params^*$ as 
\begin{align}
    \sigma(\bm{x}_0, \bm{x}_0^{[i]}) \approx \params^{*\top}[\gobs(\bm{x}_0, \bm{x}_0^{[i]}) -\gobs(\bm{x}_0^{[i]},\bm{x}_0)]/2 = \sum_{i \ne j} \theta^*_{ij} g_{ij}(\bm{x}_0, \bm{x}_0^{[i]}),\label{eq:appfluctep}
\end{align}
Here we used that the off-diagonal observables $g_{ij}$ for $i\ne j$ are antisymmetric and the diagonal observables $g_{ii}$ are symmetric. 
Plugging the observables $\gobs$ from Eq.~\eqref{eq:smobs} into Eq.~\eqref{eq:appfluctep} gives
\begin{align}
      \sigma(\bm{x}_0, \bm{x}_0^{[i]}) \approx  \sum_{j:j\ne i} \theta^*_{ij}  (x_{i,0}^{[i]}-x_{i,0})x_{j,0}=-2\sum_{j:j\ne i} \theta^*_{ij}  x_{i,0} x_{j,0}\,.
    \label{eq:approxapptraj}
\end{align}

Using this expression, we may  write the sum~\eqref{eq:smsumAA} in terms of our inferred parameters as
\begin{align}
&\qquad\qquad\;\sigma(\bm{x}_0,\bm{x}_0^{[i]})\;+\;\sigma(\bm{x}_0^{[i]},\bm{x}_0^{[i,k]})\;\;+\;\;\sigma(\bm{x}^{[i,k]},\bm{x}_0^{[k]})\;\;\;\;+\;\;\;\;\sigma(\bm{x}_0^{[k]},\bm{x})\nonumber\\
& \approx-2\left[\sum_{j: j \ne i} \theta^{*}_{ij} x_{i,0} x_{j,0} + \sum_{j: j \ne k} \theta^{*}_{kj} x_{k,0}^{[i]} x_{j,0}^{[i]} + \sum_{j: j \ne i} \theta^{*}_{ij} x_{i,0}^{[i,k]} x_{j,0}^{[i,k]} + \sum_{j: j \ne k} \theta^{*}_{kj} x_{k,0}^{[k]} x_{j,0}^{[k]} \right]\nonumber\\
&= -2\left[\sum_{j: j \ne i} \theta^{*}_{ij} x_{i,0} x_{j,0} + \sum_{j: j \ne k} \theta^{*}_{kj} x_{k,0} x_{j,0}^{[i]} -\sum_{j: j \ne i} \theta^{*}_{ij} x_{i,0} x_{j,0}^{[i,k]} - \sum_{j: j \ne k} \theta^{*}_{kj} x_{k,0} x_{j,0}^{[k]}\right]\nonumber\\
&= -2\left[\sum_{j: j \ne i} \theta^*_{ij} x_{i,0} x_{j,0} + \sum_{j: j \ne k} \theta^*_{kj} x_{k,0} x_{j,0}^{[i]} -\sum_{j: j \ne i} \theta^*_{ij} x_{i,0} x_{j,0}^{[k]} - \sum_{j: j \ne k} \theta^*_{kj} x_{k,0} x_{j,0}\right]\nonumber\\
&= -4(\theta_{ik}^{*}-\theta_{ki}^{*})x_{i,0}x_{k,0} . %
\label{eq:fluctWW}
\end{align}
By comparing the two expressions, we have shown that for $i \ne k$,
\begin{equation}
    \beta(w_{ik}-w_{ki})\approx \theta_{ik}^{*}-\theta_{ki}^{*}\,.
    \label{eq:app_w-theta}
\end{equation}

\subsection{Optimization-free inference of the EP phase transition in the asymmetric Sherrington-Kirkpatrick model}

\begin{figure}[h]
    \centering
    \begin{tikzpicture}[remember picture]
        \node[anchor=north west, inner sep=0] at (0.1,0) {\includegraphics[width=6cm]{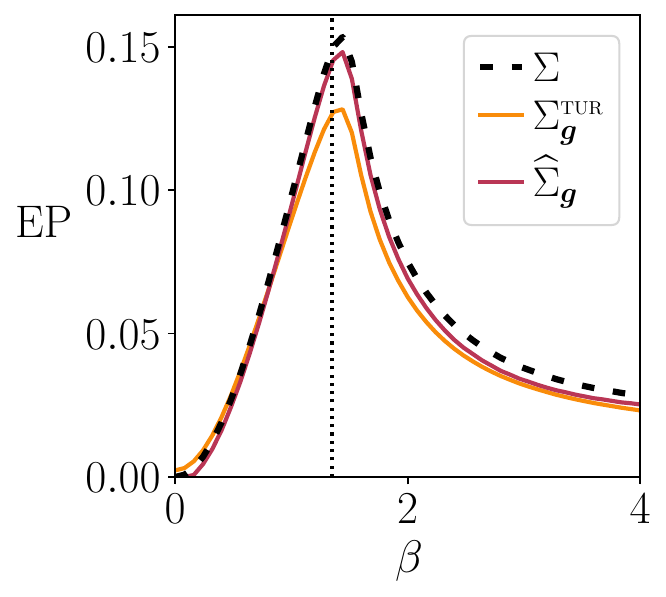}};
        \node[anchor=north west, inner sep=0] at (8.3,0) {\includegraphics[width=6cm]{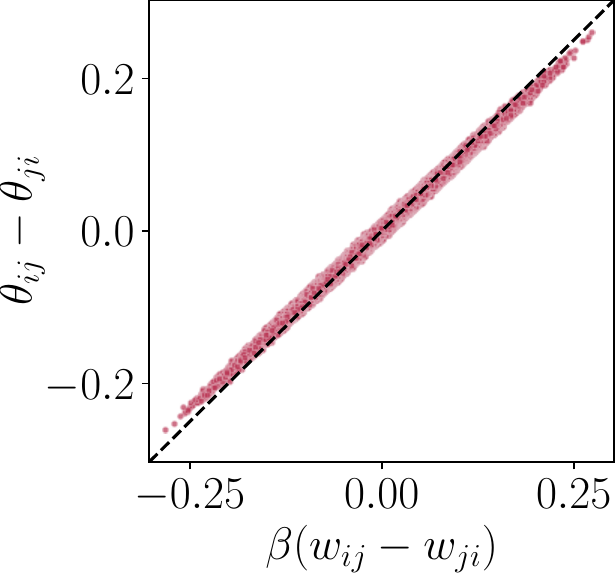}};
        \node[anchor=north west] at (0.0,0.) {\small\textbf{(a)}};
        \node[anchor=north west] at (8.2,0.) {\small\textbf{(b)}};
    \end{tikzpicture}
    \caption{\textbf{(a)} EP estimators $\epGtur$ and $\GepG$ for the fully connected asymmetric Sherrington–Kirkpatrick model. %
    \textbf{(b)} Inferred parameters corresponding to  $\GepG$ at the critical point $\beta \approx 1.3485$. Despite the presence of a critical point (dotted line in \textbf{(a)}), where standard inference methods struggle \cite{aguilera2021unifying}, $\GepG$ accurately recovers the full EP.}
    \label{fig:S-spin-models}
\end{figure}

To test the robustness of our estimator $\GepG$, we also evaluated it on a dense variant of the spin model. Specifically, we considered the case $k = N - 1$, where each spin is connected to all others. In this fully connected regime, the network converges to the asymmetric Sherrington–Kirkpatrick (SK) model, a classical mean-field model with quenched disorder and Gaussian couplings. We use a model with couplings $w_{ij} = w_0/N + \gamma z_{ij} / \sqrt{N}$, with $z_{ij} \sim \mathcal{N}(0, 1)$.

For $w_0=1$ and $\gamma=1/2$, this model has a known ordered phase in the large-$\beta$ limit, a second-order 
order-disorder critical point at $\beta\approx 1.3485$ where EP peaks~\cite{aguilera2023nonequilibrium}.  
Near the critical point, classical mean-field methods such as the TAP approximation fail in nonequilibrium MaxEnt parameter inference, leading to unreliable estimates of EP \cite{aguilera2021unifying}. 

Figure~\ref{fig:S-spin-models} illustrates the results for the asymmetric SK model for $w_0=1,\gamma=1/2$. We repeated our analysis using this dense coupling structure and found that $\GepG$ (with similar holdout methods as above) successfully recovered the full entropy production (EP). Although we do not plot it here, we verified that $\epG$ gives the same values. This demonstrates that our method performs reliably in disordered systems with approximately Gaussian statistics. %

\section{LINEAR LANGEVIN PROCESS}
\label{sec:ou}

To explore the applications for continuous-valued systems, we benchmark our estimator on an $N$-dimensional linear Langevin process
\begin{align}
\dot{\bm x}(t)= \bm A \bm x(t) + \sqrt{2} \bm B \bm\xi(t),
\end{align}
with drift matrix $\bm A$, noise matrix $\bm B$, and standard white noise $\bm\xi(t)$.
The diffusion matrix is given by $\bm D =  \bm B \bm B^\top$. Stationarity requires $\mathrm{Re}\,\lambda(A) < 0$ for all eigenvalues $\lambda(A)$.

We define time-lagged covariances as $\bm S_{ts} := \langle \bm x(t) \bm{x}(s)^\top\rangle$. The steady-state same-time covariance solves the continuous Lyapunov equation
\begin{align}
\bm  A \bm S_{tt} + \bm S_{tt}  \bm A^\top + 2\bm D= \bm{0}.
\label{eq:lyap}
\end{align} 
The lag-$t$ covariance follows from linearity:
\begin{align}
S_{t0}=
\begin{cases}
e^{\bm A t}\, \bm S_{00}, & t \ge 0,\\
\bm S_{00}\, e^{\bm A^\top (-t)}, & t < 0.
\end{cases}
\label{eq:lagcov}
\end{align}

For any $t>0$, the joint $p_{0,t}(\bm x(0), \bm x(t))$ in steady state is a zero-mean Gaussian with block covariance
\begin{align}
\bm  S = 
\begin{bmatrix}
\bm S_{00} & \bm S_{t0}^\top \\
\bm S_{t0} & \bm S_{tt}
\end{bmatrix},
\end{align}
where at steady state $\bm S_{tt}=\bm S_{00}$.

The time-reversed joint $\tilde p_{0,t}(\bm x(0),\bm x(t))= p_{0,t}(\bm x(t),\bm x(0))$ has covariance
\begin{align}
\tilde {\bm S}=
\begin{bmatrix}
\bm S_{tt} & \bm S_{t0} \\
\bm S_{t0}^\top  & \bm S_{00}
\end{bmatrix}.
\end{align}

Then, the (discrete-time, lag-$t$) entropy production in the steady state is given by the KL divergence between these two Gaussian distributions,
\begin{align}
\ep &= D[p(\bm x(0),\bm x(t))\,\Vert\, \tilde p(\bm x(0),\bm x(t))] 
\nonumber \\ &=  \frac{1}{2} \mathrm{Tr} \big[ \tilde {\bm S}^{-1} \bm S \big] - N ,
\label{eq:gauss-kl}
\end{align}
which serves as ground truth for our estimators.

\begin{figure}[t]
    \centering
    \includegraphics[width=0.6\linewidth]{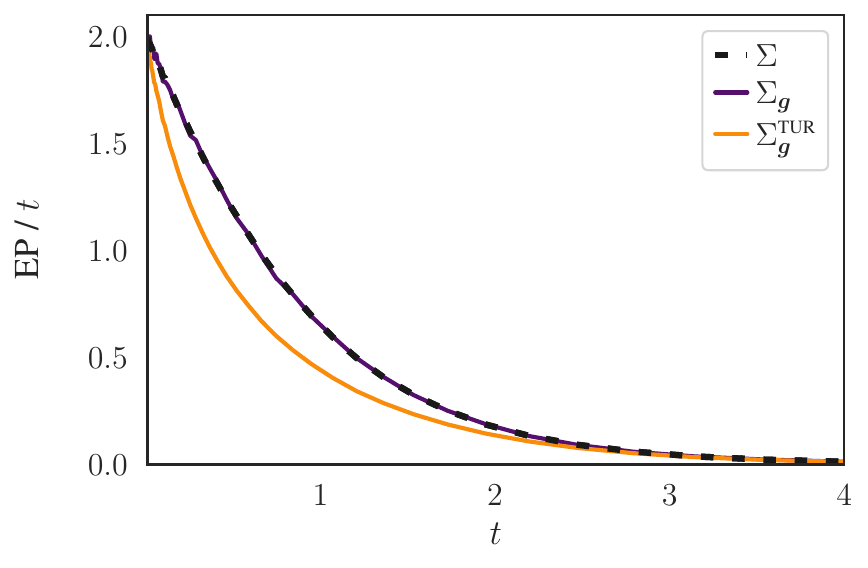}
    \caption{\textbf{Linear Langevin model.} EP per lag time $\ep(t)/t$ versus our estimators for an $N=10$ system with $T=4 \times 10^6$ samples per lag, and lags distributed logarithmically. Our estimator %
    $\epG$ is tight, while the MTUR estimator is looser for intermediate $t$.}
    \label{fig:ou}
\end{figure}

We take $\bm A=- \bm I+ k \bm W$ with $W_{ij}\sim \mathcal N(0,\sigma^2)$ and $\bm B=\bm I$. We choose $k=0.2$, ensuring $\mathrm{Re}\,\lambda(\bm A)<0$ for all eigenvalues.

As our observables, we use cross-correlations and changes of same-time correlations:
\begin{align}
    \gobs = (\gobs^\mathrm{cross}, \gobs^\mathrm{same}),\qquad \qquad \begin{aligned}
    g_{ij}^\mathrm{cross} &:= x_{i}(t)x_{j}(0)-x_{i}(0)x_{j}(t) \quad (i<j),
\\
g_{ij}^\mathrm{same} &:= x_{i}(t)x_{j}(t)-x_{i}(0)x_{j}(0) \quad (i<j).
    \end{aligned}
\end{align}
For different values of $t$, we draw $4\times 10^6$ i.i.d. pairs $(\bm x_0,\bm x_t)$ as above, split them into train/validation/test (70\%/20\%/10\%), and optimize $\epG$ with early stopping as detailed in Sec.~\ref{app:optimization} (gradient ascent with Barzilai–Borwein steps). We report test-set values.

Figure~\ref{fig:ou} shows $\ep/t$  together with $\epG/t$ and $\epGtur/t$ for a representative $N=10$ system. Our  estimator $\epG$ accurately tracks the exact Gaussian EP across different timescales. The MTUR is valid in the short-time and long-time limits, but looser for intermediate values. The looseness of the MTUR arises because our observables involve products of Gaussians, and thus are not themselves Gaussian distributed.

Our results confirm that our estimator recovers the correct EP for linear Langevin systems. Moreover, it can do this using only %
expected cross-covariances, with no need to reconstruct probability flows or drift.

\section{NEUROPIXELS DATASET}

We use high-resolution spiking data obtained from the Allen Institute's Neuropixels dataset \cite{allen2022visual}. The dataset consists of multi-region neural recordings from awake mice, capturing parallel spike trains from hundreds of neurons simultaneously across brain areas.

\subsection{Data selection and preprocessing}

We use the AllenSDK Python library to download the %
data and preprocess it to obtain binned binary spike trains for populations of neurons. For each session, we follow these steps:
\begin{itemize}
    \item Session filtering: From 153 recorded sessions, we take the 103 sessions flagged as valid by default from the AllenSDK library.
    \item Unit filtering: We restrict the dataset to well-isolated units by applying standard quality control filters: signal-to-noise ratio (SNR) greater than 1, inter-spike interval (ISI) violations less than 1, and mean firing rate above 0.1 Hz. This ensures that only stable, active neurons are included.
    \item Spike binning: For each selected unit, we discretize its spike train using non-overlapping bins of 10~ms.  This results in a high-dimensional binary time series with parallel (non-multipartite) updates.
    \item Condition segmentation: Each recording session includes different behavioral and stimulus conditions, which are annotated in the Allen metadata. We define temporal masks to extract three conditions:
    \begin{enumerate}
        \item Active behavior (60 min): periods where the animal was engaged in a visual discrimination task.
        \item Passive replay (60 min): periods where the animal passively viewed natural images.
        \item Gabor stimuli (25 min): control periods where drifting gratings (Gabor patches and full-field flashes) were shown instead of images.
    \end{enumerate}
    Using the provided timestamps, we generate masks to extract the corresponding segments from the full spike train, yielding separate datasets for each condition. Under the selected bin size, each condition contains approximately 360,000 (active and passive) and 150,000 (Gabor) data points.
    \item Brain areas: Each unit is associated with a brain region via its recording channel. The dataset includes activity from visual cortical areas (VISp, VISl, VISal, VISrl, VISam and VISpm) as well as subcortical structures, such as the lateral geniculate nucleus, lateral posterior nucleus, hippocampus, and midbrain. We record these region labels alongside the spike data for the analysis shown in Fig.~\ref{fig:neuropixels}(b). 
\end{itemize}

\subsection*{Optimization details}

The Neuropixels dataset allows for parallel (non-multipartite) transitions where multiple degrees of freedom change state at once. For this reason, we cannot factorize our optimization problem into independent problems (using the technique discussed in the \emph{End Matter}), and our optimization involves a single problem with (up to) tens of thousands of parameters. Nonetheless, this problem can be solved relatively quickly using the gradient ascent method described above.

To ensure sufficient activity for reliable estimation, we only included neurons that were active in at least $2\%$ of time bins. For each population size $N$, %
we averaged results across 10 independent trials, each using a different random subset of neurons. To visualize the inferred coupling parameters in Fig.~\ref{fig:neuropixels}(b), we selected 200 neurons with the highest firing rate from an active trial.

\bibliography{references}

\clearpage